\documentclass[preprint,showpacs,preprintnumbers,amsmath,amssymb,noshowpacs]{revtex4-1}
\usepackage{graphicx}
\usepackage{dcolumn}
\usepackage{mathrsfs}
\usepackage{bm}

\begin{document}

\title{Quantum state engineering by nonlinear quantum interference}

\author{Liang Cui$^{1}$}
\thanks{These authors contributed equally to this work.}
\author{Jie Su$^{1,3}$}
\thanks{These authors contributed equally to this work.}
\author{Jiamin Li$^{1}$}
\author{Yuhong Liu$^{1}$}
\author{Xiaoying Li$^{1}$}
\email{xiaoyingli@tju.edu.cn}
\author{Z. Y. Ou$^{1,2}$}
\email{zheyuou@tju.edu.cn}
\affiliation{%
$^{1}$College of Precision Instrument and Opto-Electronics Engineering, Key Laboratory of
Opto-Electronics Information Technology, Ministry of Education, Tianjin University,
Tianjin 300072, P. R. China\\
$^{2}$Department of Physics, Indiana University-Purdue University Indianapolis, Indianapolis, IN 46202, USA\\
$^{3}$School of Electronic and Control Engineering, North China Institute of Aerospace Engineering, Langfang 065000, China
}%

\date{\today}

\begin{abstract}
Multi-photon quantum interference is the underlying principle for optical quantum information processing protocols. Indistinguishability is the key to quantum interference. Therefore, the success of many protocols in optical quantum information processing relies on the availability of photon states with a well-defined spatial and temporal mode. Photons in single spatial mode can be obtained from nonlinear processes in single-mode waveguides. For the temporal mode, the common approach is to engineer the nonlinear processes so as to achieve the required spectral properties for the generated photons. But this approach is complicated because the spectral properties and the nonlinear interaction are often intertwined through phase matching condition. In this paper, we study a different approach that separates the spectral control from nonlinear interaction, leading to versatile and precise engineering of the spectral properties of nonlinear parametric processes. The approach is based on an SU(1,1) nonlinear interferometer with a pulsed pump and a controllable linear spectral phase shift for precise engineering. We systematically analyze the important figures of merit such as modal purity and heralding efficiency in characterizing a photon state and use this analysis to investigate the feasibility of this interferometric approach. Specifically, we analyze in detail the requirement on the spectral phase engineering to optimize the figures of merit and apply numerical simulations to the nonlinear four-wave mixing process in dispersion-shifted fibers with a standard single-mode fiber as the phase control medium. Both modal purity and efficiency are improved simultaneously with this technique. Furthermore, a novel multi-stage nonlinear interferometer is proposed and shown to achieve more precise state engineering for near-ideal single-mode operation and near-unity efficiency. We also extend the study to the case of high pump power when the high gain is achieved in the four-wave mixing process for the spectral engineering of quantum entanglement in continuous variables. Our investigation provides a new approach for precisely tailoring the spectral property of quantum light sources, especially, photon pairs can be engineered to simultaneously possess the features of high purity, high collection efficiency, high brightness, and high flexibility in wavelength and bandwidth selection.
\end{abstract}
\maketitle

\section{Introduction}

Many protocols in quantum information and quantum communication were first demonstrated in optics \cite{Bennett98,Tittel01} because of the simplicity in photons and the easiness to implement them with linear optics \cite{KLM,Kok07}. This requires high quality single-photon and multi-photon sources with superior modal purity and efficiency. One approach is to produce single photon on demand \cite{spd}. Despite the constant improvement of technology that leads to high quality in photon indistinguishability of the single-photon source~\cite{pan16PRL}, this type of photon source still lacks the consistency in repeatability, that is, the quality varies from one source to another. This limits the applicability of the source. Another common approach that began from the early stage is the correlated photon pair generation from spontaneous emission of nonlinear parametric processes, which has become a popular multi-photon source ever since its discovery.  A single-photon state can be produced by heralding on the detection of one of the photon pair \cite{herald}. Because of its simplicity, this type of photon source has been used in a wide range of applications in quantum information processing (QIP).

Because of the way they are generated, the photon pairs from spontaneous parametric emission (SPE) are highly correlated in frequency and time. This, on the one hand, is highly desirable in studying quantum entanglement in frequency and time, on the other hand leads to distinguishability in time due to difference in the production time of the photon pairs and becomes troublesome for quantum interference, in particular, in the QIP protocols  involving the quantum interference among multiple sources, such as the generation of multi-photon entanglement \cite{wang-pan16} and quantum teleportation \cite{Bou97}. To tackle this problem, ultra-short pulses are used to eliminate the time uncertainty and define a proper temporal mode \cite{zuk,rar}. This effort, however, was hampered by the dispersion in nonlinear optical media due to the ultra-fast process \cite{Ou97,Grice97} and leads to even more complicated temporal modes. Ironically, to obtain a better temporal mode for the two-photon fields, it is desirable to have no frequency correlation between the photons so that each photon can have a definite temporal mode of their own \cite{Grice-PRA-2001}. This leads to the requirement of factorization of two-photon wave-function or joint spectral function (JSF) \cite{ou99}.

Efforts in acquiring a factorized JSF have been underway for quite some years ever since it was discovered that high visibility in multi-photon quantum interference relies on the factorization of the JSF \cite{ou99}. In the early days, the factorization of JSF was realized by utilizing passive filtering \cite{Bou97,zuk,rar}. However, it is well known that this method will result in a reduction of brightness. Moreover, the collection efficiency of photon pairs, which corresponds to the heralding efficiency of heralded single photons, will be significantly reduced because the filtering process will cut out photons randomly to destroy photon correlation and degrades the quality of the quantum correlated photon pairs.~\cite{Ou97,Grice97,Brida2011}. Then came the idea of engineering the source of photon pairs to achieve factorization without filtering. Over the years, many techniques have been deployed to directly engineer the JSF into a factorized form. They include the employment of photonic grating for active temporal mode shaping \cite{raymer}, special selection of $\chi^{(2)}$-nonlinear crystals with the desired properties \cite{Mosley08}, engineering of the dispersion of nonlinear optical fiber \cite{Garay-Palmett-OE2007,xylOE,Walmsley-tailored,Cui2012NJP}, and engineering of the structure of the nonlinear photonic crystals \cite{Evans10PRL,Branczyk11OE}.

The common goal in the techniques mentioned above is to engineer the JSF by manipulating the linear spectral properties of the nonlinear media to achieve an un-correlated and near factorized JSF without passive filtering. The key parameters for a successful engineering are the high modal purity and the good collection or heralding efficiency while maintaining a high photon pair production rate. While most have achieved the aforementioned goals to some extent, many are limited to specific wavelengths of operation due to strict requirement on dispersion and are therefore lack of tunability.

Two factors need to be considered in the engineering of the JSF: (1) dispersion of the media for tailoring the spectral shape of JSF and (2) phase matching for achieving efficient nonlinear interaction. Most of the schemes implemented so far for quantum state engineering have the two aspects intertwined: changing one will affect the other and everything has to be just right to achieve the goals. This is why most of the schemes are lack of tunability.

In this paper, we consider a totally different approach in which we separate the nonlinear gain control and dispersion engineering by the method of SU(1,1)-type nonlinear quantum interference~\cite{Chekhoa16}. The SU(1,1)-type nonlinear interferometer (NLI), first proposed by Yurke et al. \cite{Yurke86} and recently realized experimentally~\cite{jing11, Hud14},  is analogous to a conventional Mach-Zehnder interferometer (MZI) but with the two splitting mirrors being substituted by two nonlinear media. Originally designed to achieve the Heisenberg limit in precision phase measurement, this type of NLI has found applications in quantum interferometry beyond standard quantum limit~\cite{Hud14}, imaging with undetected photons~\cite{Lemos14},  and infrared spectroscopy~\cite{Dmitry16}, and has been realized with atoms in a Bose-Einstein condensate \cite{BEC}, phonons in an opto-mechanical system \cite{OM}, microwaves in low noise RF amplifiers \cite{fl}, and a combined atom-photon system in hybrid atom-light interferometers \cite{Chen15}. It was further suggested and demonstrated that SU(1,1) interferometers could be used to modify the spectra of the output fields \cite{tatia, Chekhova16, Sharapova18PRA, Huo19arx}. Our group recently used the SU(1,1) interferometer and achieved precise engineering of the joint spectral function of photon pairs with versatile selection of wavelengths \cite{Su19OE, LJM20arx}.
Following this idea, here in this paper, we analyze theoretically in a systematic way the method of reshaping the JSF of photon pairs based on the SU(1,1)-type nonlinear interferometer, in which the phase matching of the parametric processes is controlled by the nonlinear media whereas the spectral shaping is achieved via dispersive phase control of the interferometer. With the roles of phase matching and spectral reshaping separated, we study in detail the requirement on the dispersive property of the linear media to achieve versatile engineering of the JSF and accordingly
design the required dispersive phase with a programmable optical filter commonly employed in ultra-fast pulse shaping \cite{Weiner2000,Hu2017}. Better control and finer engineering of the JSF can also be achieved with a novel multi-stage nonlinear interference scheme for the production of higher quality two-photon state. We study this multi-stage NLI under various conditions for optimum engineering of the JSF. We further investigate the high gain regime for the parametric processes where quantum entanglement in continuous variables is possible. The involvement of dispersive media in the interference process leads to active spectral filtering, which, different from passive filtering with regular filters, maintains the original high collection efficiency for good photon heralding efficiency and keeps in the meantime a good modal purity with high brightness, all desirable in many quantum information protocols.

The rest of the paper is organized as follows. We first lay the groundwork for quantum state engineering in Sect. II with a characterization of multi-mode two-photon state from spontaneous parametric emission (SPE) processes by defining some key parameters such as state purity and heralding efficiency. Then, we introduce the SU(1,1)-type NLI in Sect. III for the engineering of JSF and apply it to an optical fiber system and demonstrate the improvement of the key parameters by the new scheme. To make a better control and finer engineering, we introduce the techniques of programmable optical filtering and multi-stage interference in Sect. IV. We extend the analysis of the NLI to the high gain regime of the parametric processes in Sect. V. Finally, we conclude with a summary and discussion in Sect. VI.

\section{Generation and characterization of two-photon states and heralded single-photon states by spontaneous parametric processes}

\subsection{Two-photon states and Schmidt mode decomposition}

Two-photon states are usually generated in the signal and idler field of spontaneous parametric emission (SPE) processes through nonlinear interactions of three- or four-wave mixing with one or two strong pump fields. When the pump power is relatively low, the dominating interaction leads to two-photon generation. If the spatial modes are well-defined, as in optical fiber, we can use one-dimensional description for the generated signal and idler fields and the output quantum state takes the form of
\begin{eqnarray}\label{TPS}
|\Psi\rangle \approx |vac\rangle + G|\Psi_2\rangle
\end{eqnarray}
with the two-photon state term
\begin{eqnarray}\label{TPS2}
|\Psi_2\rangle = \int d \omega_s d\omega_iF(\omega_s,\omega_i)\hat a_s ^{\dag}(\omega_s)\hat a_i^{\dag}(\omega_i)|vac\rangle,
\end{eqnarray}
where $\hat a_s^{\dag}(\omega_s)$ and $\hat a_i^{\dag}(\omega_i)$ are the creation operators of the signal and idler fields at $\omega_s$ and $\omega_i$, respectively. The coefficient $G$ determined by the pump power and effective length of nonlinear interaction is proportional to the gain of SPE. The JSF $F(\omega_s,\omega_i)$ is normalized as
\begin{eqnarray}\label{JSF-NOR}
\int d\omega_sd\omega_i |F(\omega_s,\omega_i)|^2 = 1
\end{eqnarray}
and can be expressed via singular mode decomposition method as Schmidt mode expansion \cite{law,guo-OE16}:
\begin{eqnarray}\label{JSF}
F(\omega_s,\omega_i)= \sum_k r_k \psi_k(\omega_s) \phi_k(\omega_i)
\end{eqnarray}
with mode expansion coefficients $r_k\ge0$ $(k=1,2,...)$, $\sum_k r^2_k=1$, and two sets of orthonormal functions $\{\psi_k(\omega_s), \phi_k(\omega_i)\}$ satisfying
\begin{eqnarray}\label{ornorm}
\int d\omega_s\psi^*_k(\omega_s) \psi_{k'}(\omega_s) = \delta_{kk'} = \int d\omega_i\phi_k^*(\omega_i) \phi_{k'}(\omega_i).
\end{eqnarray}

With mode decomposition in Eq.(\ref{JSF}), the state in Eq.(\ref{TPS}) can be rewritten as
\begin{eqnarray}\label{TPS-rk}
|\Psi\rangle \approx |vac\rangle + G|\Psi_2\rangle \approx |vac\rangle +G\sum_k r_k \hat A_k ^{\dag}\hat B_k^{\dag}|vac\rangle =|vac\rangle+G\sum_k r_k |1_k\rangle_s|1_k\rangle_i,
\end{eqnarray}
where operators
\begin{eqnarray}\label{AB-k}
\hat A_k ^{\dag} \equiv \int d\omega_s \psi_k(\omega_s)\hat a_s^{\dag}(\omega_s),~~\hat B_k ^{\dag} \equiv \int d\omega_i \phi_k(\omega_i)\hat a_i^{\dag}(\omega_i)
\end{eqnarray}
define single temporal modes for the signal and idler fields, respectively. $|1_k\rangle_s \equiv \hat A_k ^{\dag}|vac\rangle$, $|1_k\rangle_i \equiv \hat B_k ^{\dag}|vac\rangle$ are the single-photon states in those temporal modes \cite{guo-OE16}. The way in which $|\Psi\rangle$ is expressed in terms of the temporal modes in Eq.(\ref{TPS-rk}) indicates that it is a multi-mode two-photon state and is in the form of high-dimensional entanglement \cite{law,Morandotti17}. The Schmidt mode number $K$ is defined through the coefficients $r_k$ by
\begin{eqnarray}\label{K}
K \equiv 1\bigg / \sum_k r_k^4.
\end{eqnarray}
Take, for example, the case of $M$ modes with equal weight: $r_k^2 = 1/M$ ($k=1,2,...,M$) but $r_k=0$ for other $k$. We have from Eq.(\ref{K}) $K= 1/(M\times (1/M^2)) = M$, i.e., the number of modes. Hence, the Schmidt number is an approximate measure of the number of modes in the two-photon state $|\Psi_2\rangle$ in Eq. (\ref{TPS-rk}).

Experimentally, it is hard to measure the JSF and make the decomposition in Eq. (\ref{JSF}). Thus, it is impractical to use Eq.(\ref{K}) to obtain the mode number. On the other hand, it has been shown that the measurable quantity $g^{(2)}_{s(i)}$, i.e., the normalized intensity correlation of the individual signal (idler) field alone, which comes from the higher-order terms in spontaneous parametric process (see later in Eqs. (\ref{TFPS}) and (\ref{FPS})), can be expressed in terms of the Schmidt number as \cite{ou99,LiuNN-OE16}
\begin{eqnarray}\label{g2}
g^{(2)}_{s(i)}  \equiv \frac{\int dt_1dt_2 \langle :\hat I_{s(i)}(t_1)\hat I_{s(i)}(t_2):\rangle}
{\big[\int dt \langle\hat I_{s(i)}(t)\rangle \big]^2} = 1+{\frac{\cal E}{\cal A}}  = 1 +  \sum_k r_k^4 = 1+ \frac{1}{K},
\end{eqnarray}
where $\hat I_{s(i)}(t) = \hat E_{s(i)}^{\dag}(t)\hat E_{s(i)}(t)$ with $\hat E_{s(i)}(t)=\frac{1}{\sqrt{2\pi}}\int d\omega \hat a_{s(i)}(\omega) e^{-j\omega t}$ being the electric field operator of the signal (idler) field and
\begin{eqnarray}\label{EA}
{\cal E}&\equiv &  \int d\omega_sd\omega_i d\omega_s'd\omega_i' F(\omega_s,\omega_i) F(\omega_s',\omega_i')F^*(\omega_s,\omega_i') F^*(\omega_s',\omega_i)= \sum_k r_k^4\cr
{\cal A}&\equiv & \int d\omega_sd\omega_i d\omega_s'd\omega_i' |F(\omega_s,\omega_i)|^2 |F(\omega_s',\omega_i')|^2=1.
\end{eqnarray}
Thus, the measurement of $g^{(2)}_{s(i)}$ will lead to $K$ or the number of modes of the two-photon fields. $g^{(2)}_{s(i)}=2$ or $K=1$ will be an indication for single-mode operation.

The actual function of the JSF $F(\omega_s,\omega_i)$ depends on the nonlinear processes and can be engineered accordingly for various quantum information processing tasks. One of the important tasks is to produce a transform-limited single-photon state by heralding on the detection of one of the correlated photon pair, say, the idler. So, before going to the specific form of $F(\omega_s,\omega_i)$, let us first examine in the following some key parameters such as the state purity and heralding efficiency for the characterization of the heralded single-photon state.

\subsection{Heralded single-photon state and its purity}

The heralding process is a quantum projection in the form of a detection of the idler photon at time $t$, leading to the un-normalized heralded state as
\begin{eqnarray}\label{HSPS}
|\Psi_1(t)\rangle =\hat E_i(t) |\Psi\rangle ,
\end{eqnarray}
Substituting Eq.(\ref{TPS}) into the above, we have
\begin{eqnarray}\label{HSPS2}
|\Psi_1(t)\rangle &=& \frac{G}{ \sqrt{2\pi}} \int d\omega_sd\omega_id\omega \hat a_i(\omega) e^{-j\omega t}F(\omega_s,\omega_i)\hat a_s ^{\dag}(\omega_s)\hat a_i^{\dag}(\omega_i)|vac\rangle \cr &=& \frac{G}{ \sqrt{2\pi}}  \int d\omega_sd\omega_i e^{-j\omega_i t}F(\omega_s,\omega_i)\hat a_s ^{\dag}(\omega_s)|vac\rangle,
\end{eqnarray}
where we used the commutation relation $[\hat a_i(\omega),\hat a_i^{\dag}(\omega_i)] =\delta(\omega-\omega_i)$.
If the detection process does not have a good time resolution, especially in the case of two-photon states produced by ultra-fast pulses, the heralded state is a mixed state with average over all time:
\begin{eqnarray}\label{HSPS3}
\hat \rho_1 &=& \int dt |\Psi_1(t)\rangle \langle \Psi_1(t)|\cr
&=& G^2 \int d\omega_i d\omega_sd\omega_s'F(\omega_s,\omega_i)F^*(\omega_s',\omega_i)\hat a_s^{\dag}(\omega_s)|vac\rangle \langle vac |a_s(\omega_s'),
\end{eqnarray}
where we used the relation $(1/2\pi)\int d\omega e^{j\omega t} =\delta(\omega)$. Notice that the density operator in Eq.(\ref{HSPS3}) is not normalized due to state projection. With decomposition in Eq.(\ref{JSF}) and after proper normalization, we obtain
\begin{eqnarray}\label{HSPS4}
\hat \rho_1
&=& \sum_k r^2_k \int d\omega_sd\omega_s'\psi_k(\omega_s)\psi_k^*(\omega_s')\hat a_s^{\dag}(\omega_s)|vac\rangle \langle vac |\hat a_s(\omega_s')\cr
&=& \sum_k r^2_k \hat A_k^{\dag}|vac\rangle \langle vac|\hat A_k \cr
&=& \sum_k r^2_k |1_k\rangle\langle 1_k| ,
\end{eqnarray}
where we used the orthonormal relation in Eq.(\ref{ornorm}) for $\phi_k(\omega_i)$, and $|1_k\rangle \equiv \hat A_k^{\dag}|vac\rangle$ is a single-photon state in a single temporal mode $k$ defined by $\hat A_k^{\dag} \equiv \int d\omega_s \psi_k(\omega_s) \hat a_s^{\dag}(\omega_s)$.
Eq.(\ref{HSPS4}) describes a mixed multi-mode single-photon state with a state purity of
\begin{eqnarray}\label{purity}
\gamma_P \equiv  {\rm Tr}\hat \rho_1^2 = \sum_k r^4_k = 1- \sum_k r^2_k(1-r^2_k) \le 1,
\end{eqnarray}
where we used $\sum_k r^2_k=1$ and the equal sign stands only for the single-mode case of $r_1 =1, r_k=0$ $(k\ne 1)$. Note that  we have $\gamma_P = 1/K$ from Eq.(\ref{K}). So, the non-unit purity is because of the multi-mode nature of the two-photon state in Eq.(\ref{TPS}), as expressed in the mode decomposition in Eq.(\ref{JSF}). The single-mode case of $r_1=1$ corresponds to a factorized JSF: $F(\omega_s,\omega_i)=  \psi_1(\omega_s) \phi_1(\omega_i)$ and a purity equal to 1. But non-factorized JSFs will lead to a multi-mode situation with $r_1<1$ or heralded single photon state with purity less than 1.

\subsection{Effects of passive optical filtering}

Almost all experiment involves optical filtering to discriminate against background light. While the use of passive optical filtering is necessary in experiment, its role on the properties of the filtered photon pairs are mixed. On the one hand, it can reshape the JSF to make it more factorized and improve the mode structure. On the other hand, it destroys the photon correlation between the signal and the idler fields by deleting one of the photons and leads to poor collection and heralding efficiencies, as we will see later. So, we next examine the property of the generated signal (idler) field passing through passive optical filters, which can be modeled as frequency-dependent beam splitter with amplitude transmissivity $f_{s(i)}(\omega_{s(i)})$ and reflectivity $r_{s(i)}(\omega_{s(i)})$ ($[f_{s(i)}(\omega_{s(i)})]^2+[r_{s(i)}(\omega_{s(i)})]^2 = 1$). Then the state in Eq.(\ref{TPS}) is changed to
\begin{eqnarray}\label{TPSf}
|\bar \Psi\rangle &\approx& |vac\rangle + G \int d \omega_s d\omega_iF(\omega_s,\omega_i)[f_s(\omega_s)\hat a_s ^{\dag}(\omega_s)+ r_s(\omega_s)\hat a_{sv} ^{\dag}(\omega_s)]\cr
&&\hskip 2in \times [f_i(\omega_i)\hat a_i^{\dag}(\omega_i)+ r_i(\omega_i)\hat a_{iv} ^{\dag}(\omega_i)]|vac\rangle,
\end{eqnarray}
where $\hat a_{sv}^{\dag}$ and $\hat a_{iv}^{\dag}$ denote the modes that the filters reject and are replaced by vacuum.
The un-normalized projected state after heralding is then
\begin{eqnarray}\label{HSPSf}
|\bar \Psi_1(t)\rangle &=& \frac{G}{ \sqrt{2\pi}}  \int d\omega_sd\omega_i e^{-j\omega_i t}F(\omega_s,\omega_i)f_i(\omega_i)[f_s(\omega_s)\hat a_s ^{\dag}(\omega_s)+ r_s(\omega_s)\hat a_{sv} ^{\dag}(\omega_s)]|vac\rangle.
\end{eqnarray}
The heralded photon state, after time integral similar to Eq.(\ref{HSPS3}), becomes
\begin{eqnarray}\label{HSPSf2}
{\hat {\bar \rho}}_1
&=& G^2\int d\omega_i d\omega_sd\omega_s'F(\omega_s,\omega_i)F^*(\omega_s',\omega_i) f_i^2(\omega_i)  \cr && ~~\times \big[f_s(\omega_s)f_s(\omega_s')|1_s(\omega_s)\rangle \langle 1_s(\omega_s') | + r_s(\omega_s)r_s(\omega_s')|1_{sv}(\omega_s)\rangle \langle  1_{sv}(\omega_s')| \cr && ~~~~~~ +f_s(\omega_s)r_s(\omega_s')|1_s(\omega_s)\rangle \langle 1_{sv}(\omega_s') | + r_s(\omega_s)f_s(\omega_s')|1_{sv}(\omega_s)\rangle \langle  1_{s}(\omega_s')|\big].
\end{eqnarray}
Normalization requires the evaluation of the trace of the density operator above:
\begin{eqnarray}\label{HSPSf2-Tr}
{\rm Tr}{\hat {\bar \rho}}_1
=G^2\int d\omega_sd\omega_i |F(\omega_s,\omega_i) f_i(\omega_i)|^2 = G^2\bar {\cal A}_i^{1/2},
\end{eqnarray}
where $\bar {\cal A}_i \equiv \big[\int d\omega_sd\omega_i |F(\omega_s,\omega_i) f_i(\omega_i)|^2\big]^2$ is similar to that in Eq.(\ref{EA})  but with $F(\omega_s,\omega_i)$ replaced by $F(\omega_s,\omega_i) f_i(\omega_i)$.
After tracing out the filter-rejected states $|1_{sv}(\omega)\rangle$ in Eq.(\ref{HSPSf2}) and proper normalization, we arrive at
\begin{eqnarray}\label{HSPSf3}
{{\hat {\bar \rho}}_1}' ={\rm Tr}_{sv}{\hat {\bar \rho}}_1/{\rm Tr}{\hat {\bar \rho}}_1
&=& T \sum_k \bar r_k^2 |\bar 1_k\rangle\langle \bar 1_k| + R|vac\rangle\langle vac|,
\end{eqnarray}
with
\begin{eqnarray}\label{TR}
T &\equiv &\frac{\int d\omega_s d\omega_i|F(\omega_s,\omega_i)|^2 f_s^2(\omega_s) f_i^2(\omega_i)}{\int d\omega_s d\omega_i|F(\omega_s,\omega_i)|^2 f_i^2(\omega_i)}\cr
R &\equiv &\frac{\int d\omega_s d\omega_i|F(\omega_s,\omega_i)|^2 r_s^2(\omega_s) f_i^2(\omega_i)}{\int d\omega_s d\omega_i|F(\omega_s,\omega_i)|^2 f_i^2(\omega_i)},
\end{eqnarray}
where $\bar r_k$ and $|\bar 1_k\rangle = \hat{\bar A}_k^{\dag}|vac\rangle$ are obtained by Schmidt mode expansion of the filtered JSF $\bar F(\omega_s,\omega_i) \equiv F(\omega_s,\omega_i)f_s(\omega_s)f_i(\omega_i)/{\cal N}_{si}$ with normalization constant
\begin{eqnarray}\label{Norm}
{\cal N}_{si}^2 \equiv \int d\omega_s d\omega_i|F(\omega_s,\omega_i)|^2 f_s^2(\omega_s) f_i^2(\omega_i).
\end{eqnarray}
The purity of the single-photon state in Eq.(\ref{HSPSf3}) is
\begin{eqnarray}\label{purity2}
{\gamma_P}' = {\rm Tr} ({\hat {\bar \rho}}_1'  {^2}) = T^2 \sum_k \bar r_k^4 + R^2 = 1-T^2\sum_k\bar r^2_k(1-\bar r^2_k)-2TR \le 1,
\end{eqnarray}
in which the relations $T+R=1$, $T^2+R^2 =1-2TR \le 1$ and $\sum_k \bar r^2_k=1$ are applied, and the equal sign stands only if $T=1, R=0$ and $\bar r_1=1,\bar r_k=0$ $(k\ne 1)$. The reduction of the state purity comes from two sources: (i) multi-mode nature, similar to Eq.(\ref{HSPS4}), and (ii) rejection of correlated signal photons due to filtering of the modes and thus the introduction of vacuum. The latter can be understood in terms of the quantity of collection efficiency and the heralding efficiency discussed in the following.

Another key parameter in characterizing the quality of the photon pairs is the collection efficiency of photon pairs, which is defined through the single-photon detection probability and two-photon coincidence detection probability for photon pairs.
When signal and idler photons are respectively measured by two detectors, the probability  of detecting one photon in individual signal (idler) band per pulse is expressed as
\begin{eqnarray}\label{EQ_Psi}
P_{s(i)} &=&\eta_{s(i)}G^2 \int dt \langle \hat E_{s(i)}^{\dag}(t)\hat E_{s(i)}(t)\rangle_{\bar \Psi}\cr
&=&\eta_{s(i)}G^2 \int d \omega_s  d \omega_i|F(\omega_s,\omega_i)f_{s(i)}(\omega_{s(i)})|^2,
\end{eqnarray}
where $\eta_{s(i)}$ is the total detection efficiency in the signal (idler) band, and the average $\langle \hat E_{s(i)}^{\dag}(t)\hat E_{s(i)}(t)\rangle$ is over the filtered two-photon state in Eq.(\ref{TPSf}). The two-photon coincidence detection probability per pulse of a photon pair, one from the signal and the other from idler field,  is
\begin{eqnarray}\label{P_c}
P_{c} &=&\eta_{s}\eta_{i} G^2 \int dt_1d t_2 \langle \hat E_{s}^{\dag}(t_1)\hat E_{i}^{\dag}(t_2)\hat E_{i}(t_2)\hat E_{s}(t_1)\rangle_{\bar \Psi} \cr
&=&
\eta_{s}\eta_{i}G^2\int d \omega_{s}  d \omega_{i}|F(\omega_s,\omega_i)f_s(\omega_{s})f_s(\omega_{i})|^2 .
\end{eqnarray}
Accordingly, for a photon detected in the idler (signal) band, the probability of detecting its twin photon at signal (idler) band, i.e., the collection efficiency is simply the conditional probability
\begin{equation}\label{collection}
\xi_{s(i)}
\equiv \frac{P_{c}}{P_{i(s)}}
=\frac{\eta_{s(i)}\int d \omega_{s}  d \omega_{i}|F(\omega_s,\omega_i)f_s(\omega_{s})f_i(\omega_{i})|^2 }{\int d \omega_{s} d \omega_{i}|F(\omega_s,\omega_i)f_{i(s)}(\omega_{i(s)})|^2},
\end{equation}
and the probability of a photon emerging at signal (idler) band upon the detection of an idler (signal) photon, or the heralding efficiency is
\begin{equation}\label{heraldingeff}
h_{s(i)} = \xi_{s(i)}/\eta_{s(i)}
\end{equation}

Notice that from Eq.(\ref{TR}) and the above, we arrive at $T = h_s$. We thus can relate the purity in Eq.(\ref{purity2}) and the heralding efficiency in such a way. When there is no loss of photons for the signal field, i.e.,
when $\eta_{s} =1$ and $f_s(\omega_{s})\equiv 1$, the collection efficiency $\xi_s$ is unit, or $P_c = P_{s}$, a single-photon pure state can be obtained as long as the  heralding idler field is in single mode ($\bar r_1=1$).



The passive spectral filtering also affects the value of $\bar g_{s(i)}^{(2)}$ in the filtered individual signal (idler) band, which is directly related with the mode number $K$. Using Eq. (\ref{g2}) and taking the filters placed in front of detectors into account, we have
\begin{equation}\label{g2-f}
\bar g_{s(i)}^{(2)}=  1+\frac{\bar{\cal E}_{s(i)} }{ \bar{\cal A}_{s(i)}} = 1+\frac{\int {d\omega _{s(i)} d\omega _{s(i)} '\left| {\int {d\omega _{i(s)} f_{s(i)}(\omega _{s(i)} )F^* (\omega_s ,\omega _i )f_{s(i)}(\omega _{s(i)} ')F(\omega _s' ,\omega _i)} } \right|^2 }}{\left|\int {d\omega_{s} }d\omega _{i} \left| {f_{s(i)}(\omega _{s(i)} )F(\omega _s ,\omega _i )} \right|^2 \right|^2},
\end{equation}
where $\bar{\cal E}_{s(i)},\bar{\cal A}_{s(i)}$ are given in Eq.(\ref{EA}) but with the original JSF $F(\omega _s ,\omega _i)$ replaced by the one-side-filtered JSF $f_{s(i)}(\omega _{s(i)} )F(\omega _s,\omega _i)$. The dependence on only one filter function $f_{s(i)}(\omega _{s(i)} )$ is because it is measured on one side only and has nothing to do with the filter on the other side.

On the other hand, while $\bar g_{s(i)}^{(2)}$ is an experimentally measurable quantity, the Schmidt number is related to the mode coefficients $\bar r_k$ from the two-side filtered JSF $f_{s}(\omega _{s}) f_{i}(\omega _i)F(\omega _s,\omega _i)$, from which the intensity correlation function $\bar g^{(2)}$ can be calculated in Eq.(\ref{g2}) with JSF replaced by the two-side filtered JSF. Since $\bar g^{(2)}$ takes the maximum value of 2 for factorized JSF and the more filtered two-side filtered JSF tends to be more close to a factorized function than one-side-filtered JSF, we expect  $\bar g_{s(i)}^{(2)} \le \bar g^{(2)}$. Although we cannot prove this in general, it is true for the special Gaussian shaped JSF and filtering functions \cite{Yang}. So, the experimentally measurable $\bar g_{s(i)}^{(2)}$ sets a lower bound for $\bar g^{(2)}$ which is directly related to the filter-modified Schmidt number $\bar K \equiv 1/\sum_k \bar r_k^4$ or the mode property of the filtered photon pairs.

\subsection{Effects of higher order contributions from multi-pair events}

From the discussions of last section, it seems that in order to obtain high purity heralded single photons in the signal band, we only need to improve $h_s$, which can be made equal to 1 by removing the filter in the signal field, and $\bar g^{(2)}$, which can be made equal to 2 by heavily filtering the idler field. Of course, this strategy will lead to extremely small $h_i$, which does not seem to matter that much if our interest is in the signal field only. However, when high brightness of the sources is required in some of the multi-photon experiments~\cite{wang-pan16}, higher order contributions of multi-pair events are significant and must be included. But  as we will show next, low value of $h_i$ will also hamper the purity of the heralded single-photon state due to the higher photon number events such as four-photon state.

The contributions from multi-pair events will become prominent when the pump power  in spontaneous parametric processes is high in order to increase the brightness of the source. In this case, the output quantum state in Eq.(\ref{TPS}) needs to be modified to include the next order of four-photon state as \cite{ou99}
\begin{eqnarray}\label{TFPS}
|\Psi\rangle \approx |vac\rangle + G|\Psi_2\rangle + (G^2/2) |\Psi_4\rangle
\end{eqnarray}
with $|\Psi_2\rangle$ given in Eq.(\ref{TPS2}) and
\begin{eqnarray}\label{FPS}
|\Psi_4\rangle &= &|\Psi_2\rangle\otimes |\Psi_2\rangle\cr
&=&\int d \omega_s d\omega_id \omega_s' d\omega_i'F(\omega_s,\omega_i)F(\omega_s',\omega_i')\hat a_s ^{\dag}(\omega_s)\hat a_s ^{\dag}(\omega_s')\hat a_i^{\dag}(\omega_i)\hat a_i^{\dag}(\omega_i')|vac\rangle,
\end{eqnarray}
corresponding to a four-photon state due to independent two-pair generation. Using the procedure for the heralded state in Eq.(\ref{HSPSf3}) but involving more complicated derivations (see Appendix A), we find the normalized heralded state as
\begin{eqnarray}\label{HSPSf4}
{\hat {\bar \rho}}''
=  {\cal N} \bigg[T \sum_k \bar r_k^2 |\bar 1_k\rangle\langle \bar 1_k| + R|vac\rangle\langle vac| + G^2 \hat {\tilde\rho}_2'/4\bar{\cal A}_i^{1/2}\bigg]
\end{eqnarray}
with  the two-pair contribution as a two-photon state:
\begin{eqnarray}\label{rho2}
&&\hat{\tilde\rho}_2' =  \int d \omega_s d \omega_s' d\bar\omega_s d\bar\omega_s' f_s(\omega_s)f_s(\omega_s')f_s(\bar \omega_s)f_s(\bar \omega_s') \cr &&\hskip 0.3in \times \int d\omega_id\omega_i'F(\omega_s,\omega_i)F(\omega_s',\omega_i')\big[ F^*(\bar\omega_s,\omega_i)F^*(\bar\omega_s',\omega_i')+ F^*(\bar\omega_s,\omega_i')F^*(\bar\omega_s',\omega_i)\big] \cr &&\hskip 0.3in \times
[f_i^2(\omega_i)+f_i^2(\omega_i')]|1_s(\omega_s)1_s(\omega_s')\rangle \langle 1_s(\bar\omega_s)1_s(\bar\omega_s')|,
\end{eqnarray}
where $|1_s(\omega)\rangle \equiv \hat a_s^{\dag}(\omega)|vac\rangle$. ${\cal N} <1$ is the normalization factor related to $G$. The existence of the two-photon state will reduce the purity for large $G$. But the more damaging consequence is a nonzero heralded auto-intensity correlation function $\tilde g_s^{(2)}$, which is defined as
\begin{eqnarray}
\tilde g_s^{(2)}\equiv \frac{\int dt_1dt_2 \tilde \Gamma^{(2)}_s(t_1,t_2)}{
\Big[\int dt\tilde \Gamma^{(1)}_s(t)\Big]^2}
\end{eqnarray}
with
$\tilde \Gamma^{(2)}_s(t_1,t_2) \equiv  {\rm Tr}[{\hat {\bar \rho}}''\hat I_s(t_1)\hat I_s(t_2)],
\tilde \Gamma^{(1)}_s(t)  \equiv  {\rm Tr}[{\hat {\bar \rho}}''\hat I_s(t)] $. From this definition, it is obvious that $\tilde g_s^{(2)}$ is zero for the heralded state in Eq.(\ref{HSPSf3}), which is the signature property of a single-photon state. A non-zero $\tilde g_s^{(2)}$ is the contribution from the two-photon term in the heralded state in Eq. (\ref{HSPSf4}). For the state in Eq. (\ref{HSPSf4}), $\tilde g_s^{(2)}$ can be calculated through a lengthy derivation (see Appendix) to have the form of
\begin{eqnarray}\label{Hg2}
\tilde g_s^{(2)} =  \frac{2P_c}{h_s h_i} \Big(1+ \frac{\bar{\cal E}}{\bar{\cal A}}\Big),
\end{eqnarray}
where $h_{s(i)}$ is the heralding efficiency given in Eq.(\ref{heraldingeff}), $P_c$ is given in Eq.(\ref{P_c}) with $\eta_i=1=\eta_s$ and $\bar{\cal E}, \bar{\cal A}$ are given in Eq.(\ref{EA}) but with factors $f_s^2(\omega_s)$, $f_s^2(\omega_s')$, and $f_i^2(\omega_i)$ included. .

Equation (\ref{Hg2}) shows that in order to reduce $\tilde g_s^{(2)}$ for high quality heralded  single-photon state in the signal field, we need to improve the collection efficiencies in both signal and idler fields. From the discussions in Sect.IIB and above, we find that a high quality two-photon state from spontaneous parametric emission process requires high collection efficiencies in both signal and idler fields, and a high modal purity with a factorized JSF for single-mode operation. However, such strict requirements are difficult to meet from a common two-photon source, as we will see next, unless specific attention is paid to engineer the JSF.

\subsection{An example of typical two-photon sources}

To see how well the parameters in the previous sections measure up for some common sources, we next consider a specific form of JSF from spontaneous four-wave mixing (SFWM) in a single spatial mode nonlinear optical fiber \cite{LiuNN-OE16}. In the SFWM process, the photon pairs at $\omega_s$ and $\omega_i$ are created by scattering two pump photons at $\omega_p$ through the Kerr nonlinearity in fiber, thus we have the energy conservation relation $2\omega_p=\omega_s+\omega_i$.
To generate the two-photon state with well-defined time, ultrafast pulses are deployed as the pump field and we have
\begin{eqnarray}\label{JSF-fb}
F(\omega_s,\omega_i)={\cal N}_{j} \alpha(\omega_s,\omega_i)\times \kappa(\omega_s,\omega_i),
\end{eqnarray}
where ${\cal N}_{j}$ is the normalization factor to ensure the satisfaction of Eq.(\ref{JSF-NOR}),
\begin{eqnarray}
\alpha(\omega_s,\omega_i)=\exp\bigg[-\frac{(\omega_s+\omega_i-2\omega_{p0})^2}
{4\sigma_p^2}(1+jC_p)\bigg]
\end{eqnarray}
describes the Gaussian shaped pump field with the spectral width, central frequency and linear chirp of $\sigma_p$, $\omega_{p0}$, $C_p$, respectively, and
\begin{eqnarray}\label{phi}
\kappa(\omega_s,\omega_i)=\mathrm{sinc}\bigg(\frac{\Delta k L}{2}\bigg)e^{j\frac{\Delta k L}{2}}
\end{eqnarray}
is the phase matching function with
\begin{eqnarray}\label{Deltak}
\Delta k=2k(\omega_p)-k(\omega_s)-k(\omega_i)-2\gamma P_p
\end{eqnarray}
as the wave vector mismatch and $L$ denoting the length of the fiber. In Eq. (\ref{Deltak}), $k(\omega_l)$ ($l=p, s, i$) is the wave vector at $\omega_l$, $\gamma$ is the nonlinear coefficient, and $P_p$ is the peak power of pump.

After omitting the second and higher order dispersive terms in $\Delta k$, the JSF in Eq.(\ref{JSF-fb}) can be written as:
\begin{eqnarray}\label{JSF-fb-s}
F(\Omega_s,\Omega_i) = {\cal N}_{j} \exp\bigg[-\frac{(\Omega_s+\Omega_i)^2}{4\sigma_p^2}(1+jC_p)\bigg]\times\mathrm{sinc}\bigg(\frac{\Omega_s}{A}+\frac{\Omega_i}{B}\bigg)e^{j(\frac{\Omega_s}{A}+\frac{\Omega_i}{B})}
\end{eqnarray}
where $\Omega_s=\omega_s-\omega_{s0}$ and $\Omega_i=\omega_i-\omega_{i0}$ are the frequency biases of the signal and idler photons from the perfectly phase matched frequencies of the signal and idler fields, $\omega_{s0}$ and $\omega_{i0}$, respectively, and $A=2(k^{(1)}_{p0}-k^{(1)}_{s0})^{-1}L^{-1}$, $B=2(k^{(1)}_{p0}-k^{(1)}_{i0})^{-1}L^{-1}$ with $k^{(1)}_{l}=\mathrm{d} k(\omega)/\mathrm{d}\omega|_{\omega_l}$ ($l=p0, s0, i0$) are parameters depending on the linear dispersion and length of the fiber.

\begin{figure}[htb]
 \includegraphics[width=15cm]{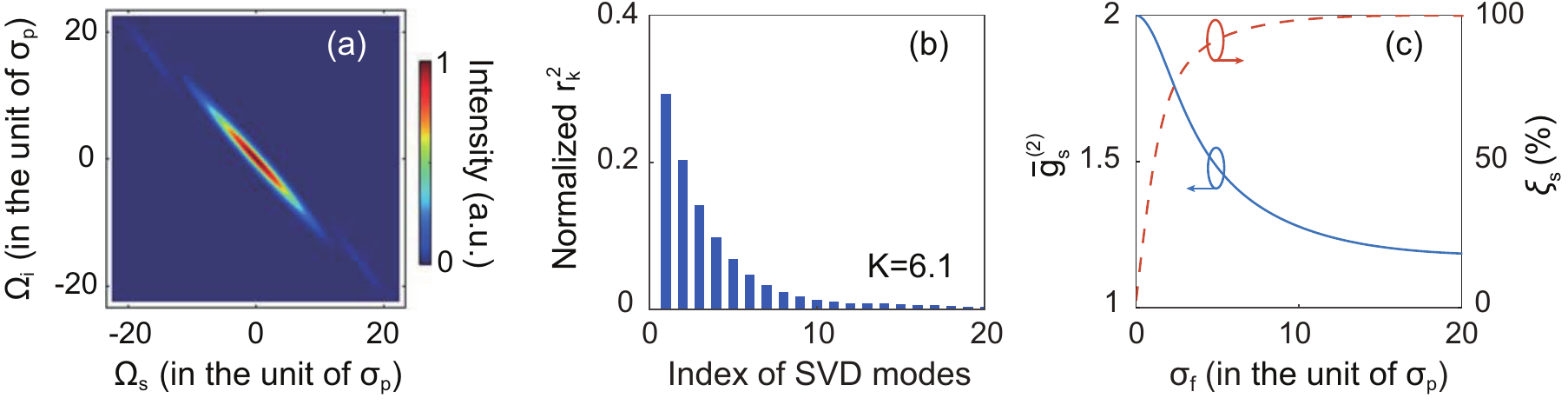}
 \caption{(a) Contour plot of the absolute square of the JSF, $|F(\Omega_s,\Omega_i)|^2$. (b) Schmidt mode expansion coefficient $r_k^2$ for $k$th mode. (c) The intensity correlation function and collection efficiency, $\bar g_s^{(2)}$ and $\xi_s$, as functions of the bandwidth $\sigma_{f}$ for filters applied to both signal and idler channels. In the simulation, the key parameters in Eq.(\ref{JSF-fb-s}) are $A=1.2 \sigma_p$, $B=1.8 \sigma_p$, and $C_p=0$.
}\label{SF}
 \end{figure}

Figure \ref{SF}(a) shows the contour plot of the JSF exhibiting anti-frequency correlation. In the calculation, the key parameters in Eq.(\ref{JSF-fb-s}) are $A=1.2 \sigma_p$, $B= 1.8 \sigma_p$, and $C_p=0$.
Note that we actually plot the absolute square of the JSF, $|F(\Omega_s,\Omega_i)|^2$, since it is directly related to the intensity of the photon pairs. The Schmidt mode expansion coefficients $r^2_k$ of the JSF is presented in Fig.\ref{SF}(b), which clearly shows the multi-mode nature with a Schmidt mode number $K=6.1$. Such a source is usually not useful for quantum information processing involving interference between independent sources.

To obtain two-photon state with a better modal purity, a straightforward method is to modify the JSF with passive optical fibers. Assuming both filters applied to the signal and idler bands are rectangular shaped with a common filter bandwidth $\sigma_f$ (see Eq. (\ref{filter}) in Sect.IIIC), we plot the intensity correlation function $\bar g_s^{(2)}$ (calculated via Eq. (\ref{g2-f})) and the collection efficiency $\xi_s$ (calculated via Eq. (\ref{collection}) with $\eta_s=1$) as functions of $\sigma_f$. As shown in Fig. \ref{SF}(c), with the decrease of bandwidth, the changing trends of $\bar g_s^{(2)}$ and $\xi_s$ are opposite, indicating that the application of narrow band filter significantly improves the modal purity but at the cost of reduced collection efficiency.

Directly engineering the JSF into a factorized form can obtain good modal purity without using filters \cite{Garay-Palmett-OE2007,xylOE,Walmsley-tailored,Cui2012NJP}. Since the pump spectral function displays the frequency anti-correlation between $\omega_s$ and $\omega_i$ due to energy conservation, to modify $F(\omega_s,\omega_i)$ into a factorable form, the dispersion of the nonlinear medium should be properly tailored to ensure the crucial condition $AB\leq0$ in Eq.(\ref{JSF-fb-s}) is satisfied \cite{Garay-Palmett-OE2007}. Note that even if it is possible, it usually only works at certain wavelengths determined by the aforementioned parameters, and there is no tunability for a pump with a fixed wavelength.

Next, we will analyze a different method of using the quantum interference of a nonlinear interferometer to engineer JSF. We will show that a non-factorable JSF, i.e., the frequency anti-correlated one for the photon pairs directly generated by one piece of nonlinear medium, can be modified to a factorable one without causing the reduction of heralding and collection efficiencies.


\section{Engineering quantum states by a two-stage nonlinear interferometer (NLI)}
We start with studying a SU(1,1)-type two stage NLI, in which the precise modal control is realized without influencing the phase
matching of the SPE in nonlinear medium. Taking the SFWM in NF as an example, we will show how the modal purity and collection efficiency can be simultaneously improved.

\subsection{Two-stage NLI}
\begin{figure}[htb]
 \includegraphics[width=10cm]{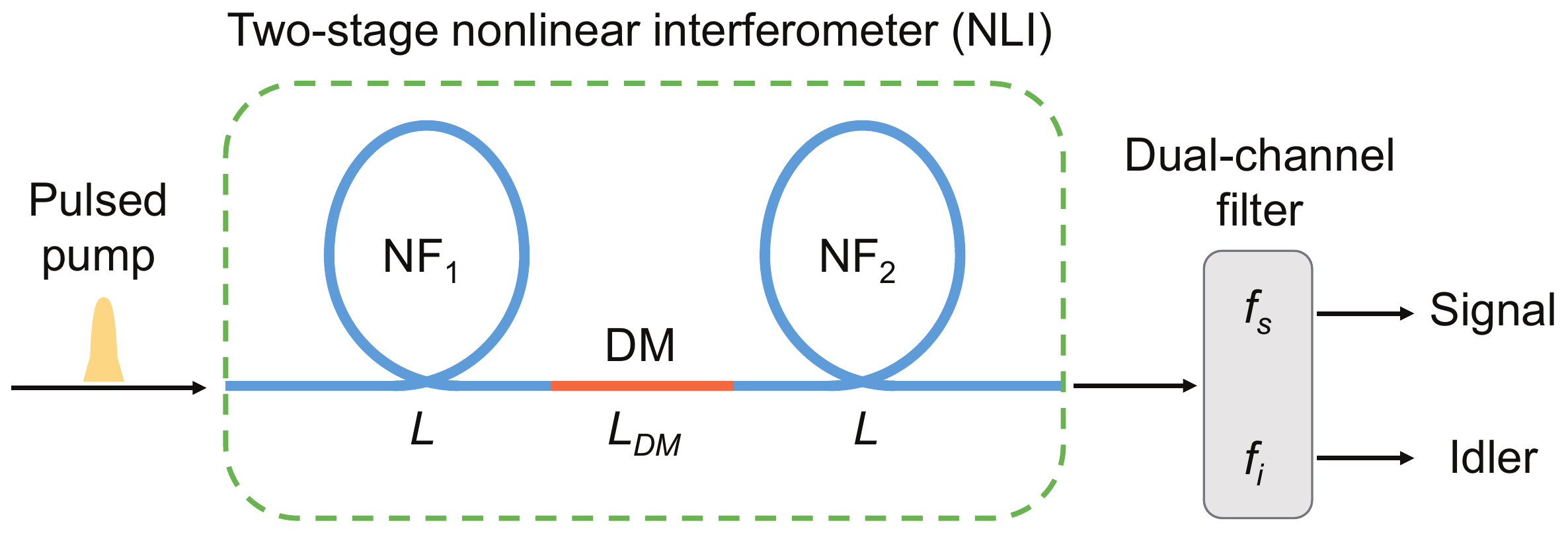}
 \caption{Schematic of the two-stage nonlinear interferometer (NLI). }\label{NLI}
 \end{figure}

Our two-stage NLI consists of two identical single-mode nonlinear fibers (NFs) with one linear dispersive medium (DM) in between, as shown in Fig. \ref{NLI}. When acting alone, each NF with length $L$ functions as a nonlinear medium of SFWM process, and the wave vector mismatch in the NF is $\Delta k$ (see Eq. (\ref{Deltak})). For a single NF being pumped by a Gaussian shaped pulse trains, the two-photon state is described by Eq. (\ref{TPS2}) with the JSF given in Eq. (\ref{JSF-fb}).
For the NLI in Fig. \ref{NLI}, quantum interference occurs between the fields produced in NF$_{1}$ and NF$_{2}$ with the phase being modulated by DM. The DM-induced phase shift, $\Delta\phi_{DM}$, is frequency(wavelength)-dependent and is a key parameter for quantum state engineering.
When the pump, signal and idler fields co-propagate through the DM, the phase shift between the three fields is then
\begin{eqnarray}\label{phase-shift}
\Delta\phi_{DM}=2\phi_{DM}(\omega_p)-\phi_{DM}(\omega_s)-\phi_{DM}(\omega_i)=\Delta k_{DM} L_{DM},
\end{eqnarray}
where $\phi_{DM}(\omega_j)$ ($j=p, s, i$) is the phase of the corresponding field after propagation,
\begin{eqnarray}\label{DeltakDM}
\Delta k_{DM}=2k_{DM}(\omega_p)-k_{DM}(\omega_s)-k_{DM}(\omega_i)
\end{eqnarray}
is the wave vector difference in the DM and $L_{DM}$ is the length of the DM.

Under the assumption of neglecting all transmission losses, the JSF of photon pairs at the output of the NLI can be calculated as
\begin{eqnarray}\label{NLI-JSF}
F_{NLI}(\omega_s,\omega_i)&=&{\cal N}_{j} \exp\bigg[-\frac{(\omega_s+\omega_i-2\omega_{p0})^2}
{4\sigma_p^2}(1+jC_p)\bigg]\times \bigg[{\rm sinc}\bigg(\frac{\Delta k L}{2}\bigg)e^{j {\frac{\Delta k L}{2}}} \cr
&&\hskip 0.3in + {\rm sinc}\bigg(\frac{\Delta k L}{2}\bigg)e^{j (\frac{\Delta k L}{2} + \Delta k L + \Delta\phi_{DM})}\bigg] \cr
&=&{\cal N}_{j} \exp\bigg[-\frac{(\omega_s+\omega_i-2\omega_{p0})^2}
{4\sigma_p^2}(1+jC_p)\bigg]{\rm sinc}\bigg(\frac{\Delta k L}{2}\bigg)  e^{j (\frac{\Delta k L}{2}+\theta)} {\rm cos}\theta,
\end{eqnarray}
with
\begin{eqnarray}\label{phase-NF2}
\theta=\frac{\Delta k L}{2}+\frac{\Delta\phi_{DM}}{2},
\end{eqnarray}
where $\cos\theta$ is the factor originated from the two-photon quantum interference. The working principle of the NLI can be explained as follows.
With the pulsed pump, both NF$_1$ and NF$_2$ can produce photon pairs, and the two photon-pair generation processes will interfere with each other. The phase difference between the two processes is the phase difference between the pump field (responsible for fields generated in NF$_2$) and the signal and idler fields launched into NF$_2$. The phase shift $\theta$ is determined by the phase mismatch in NF and the dispersion of DM, which varies with wavelength. Therefore, the overall photon-pair production rate depends on the wavelengths and this NLI functions as an active filter for photon pairs.
Usually, we have $\Delta k\rightarrow 0$ to ensure the satisfaction of phase matching, which guarantee a significant efficiency of generating photon pairs from each NF. Hence, in our scheme, $\Delta\phi_{DM}$ becomes the main term determining $\theta$ (see Eq. (\ref{phase-NF2})). This is exactly what we expect: the NFs are responsible for producing photon pairs with a certain spectrum, while the DM modifies the spectra. Therefore, in the following discussion, we will omit $\frac{\Delta k L}{2}$ in the interference factor.

In the following subsections, we will characterize the output quantum state of the NLI by substituting practical experimental parameters into the expression of JSF and simulating some key parameters discussed in Sect. II. The effect of $\Delta\phi_{DM}$ on the modification of JSF will then be visualized through the simulations.

\subsection{Phase shift induced by the DM}

From Eqs. (\ref{phase-shift}) and (\ref{DeltakDM}), one sees that the DM-induced phase shift $\Delta\phi_{DM}$ depends on the dispersion and  length of the DM. To figure out the influence of the dispersion, we start from Eq. (\ref{DeltakDM}) and expand the wave vectors at the perfectly phase-matched frequencies of the pump, signal, and idler fields,  $\omega_{p0}$, $\omega_{s0}$, and $\omega_{i0}$, respectively. After omitting the third and higher order terms, we arrive at
\begin{equation}\label{DeltakDMexpnd}
\Delta k_{DM} = \Delta k_{DM-0} + \tau_s \Omega_s + \tau_i \Omega_i + \xi_s \Omega_s^2 + \xi_i \Omega_i^2 + \xi_p \Omega_s \Omega_i,
\end{equation}
where $\Omega_{s(i)}=\omega_{s(i)}-\omega_{s0(i0)}$, and $\Delta k_{DM-0}= 2k_{DM}(\omega_{p0})-k_{DM}(\omega_{s0})-k_{DM}(\omega_{i0})$ is a constant, and
$\tau_{s(i)} = \left[ k^{(1)}_{DM}(\omega_{p0}) -  k^{(1)}_{DM}(\omega_{s0(i0)}) \right]$,
$\xi_{s(i)} = \left[ \frac{k^{(2)}_{DM}(\omega_{p0})}{4} - \frac{k^{(2)}_{DM}(\omega_{s0(i0)})}{2} \right]$, and
$\xi_{p} = \frac{k^{(2)}_{DM}(\omega_{p0})}{2}$ are parameters related to the dispersion of the DM.

\begin{figure}
	\includegraphics[width=11cm]{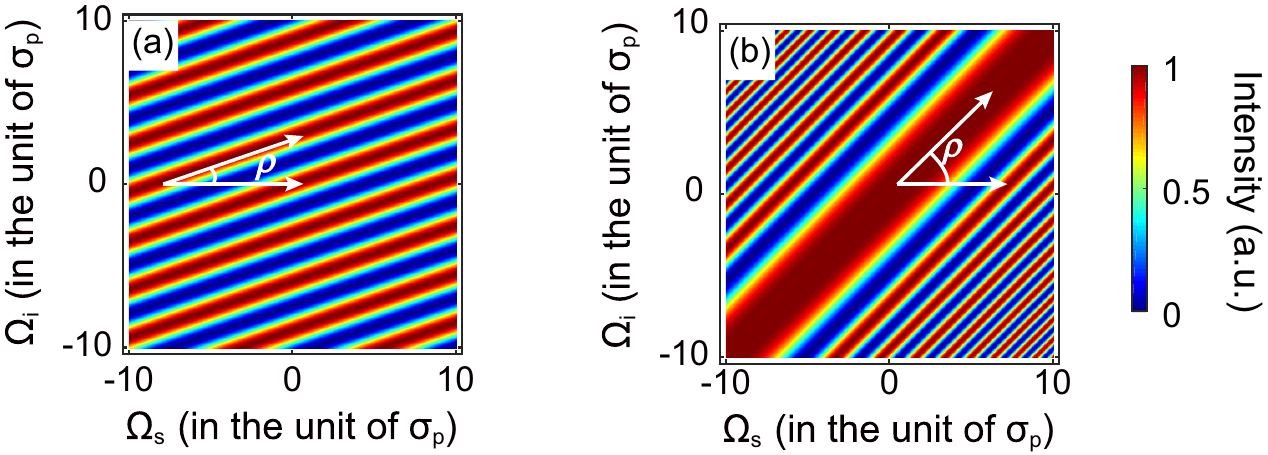}
	\caption{Contour plots of the interference factor $|{\rm cos}\theta|^2$ when the DM is a piece of silica fiber. Plot (a) shows the case of large frequency detuning between the pump and signal (idler), with $\lambda_{p0}=1053$ nm, $\lambda_{s0}=1310$ nm, $\lambda_{i0}=881$ nm, and $L_{DM} = 0.5$ m. Plot (b) shows the case of small frequency detuning, with $\lambda_{p0}=1550$ nm and $L_{DM} = 75$ m.
	} \label{Figcos}
\end{figure}

We first discuss the large frequency detuning case when the condition $|\omega_{p0}-\omega_{s0(i0)}| \ll \omega_{p0}$ is not satisfied, which means there is a significant frequency detuning (usually tens of THz) between the pump and signal (idler) fields. This kind of SFWM can be realized in photonic crystal fibers or micro/nano fibers etc. \cite{Walmsley-tailored, Cui13OL}.
In this case, the first order terms in Eq. (\ref{DeltakDMexpnd}) are dominant and the second order terms can be omitted, so we have
\begin{equation}\label{DeltakDMexpnd1}
\Delta k_{DM} = \Delta k_{DM-0} + \tau_s \Omega_s + \tau_i \Omega_i.
\end{equation}
The interference factor in Eq. (\ref{NLI-JSF}) becomes
\begin{equation}\label{eqcoslar}
{\rm cos}\theta = {\rm cos} \left[ \frac{1}{2} L_{DM} (\Delta k_{DM-0} + \tau_s \Omega_s + \tau_i \Omega_i) \right],
\end{equation}
where term $\frac{\Delta k L}{2}$ is omitted. Taking the SFWM with wavelengths of the pump, signal, and idler fields at $\lambda_{p0}=1053$ nm, $\lambda_{s0}=1310$ nm, and $\lambda_{i0}=881$ nm, respectively, as an example,
we simulate $\left| {\rm cos}\theta \right|^2$ by employing a 0.5-m-long silica fiber as the DM. Note that the angular frequency of light is related to wavelength via $\omega_l=2 \pi c \lambda_l^{-1}$ ($l=p0, s0, i0$) with $c$ denoting the speed of light in vacuum. Using the dispersion of bulk silica to approximate that of silica fiber, we find
the contour of $\left| {\rm cos}\theta \right|^2$ has an appearance of parallel stripes with equal periodicity, as shown in Fig. \ref{Figcos}(a). The orientation angle  $\rho$ of the strips is determined by the first order dispersion of the DM through
\begin{equation}\label{eqrho}
\rho = - {\rm arctan} (\frac{\tau_s}{\tau_i}).
\end{equation}
Generally, since the contour of the pump envelop always has a fixed angle of -45$^\circ$ (see Eq. (\ref{JSF-fb-s})), it is desirable to have $0 ^\circ < \rho < 90 ^\circ$ to achieve the goal of flexibly engineering the JSF. Hence, $\tau_s$ and $\tau_i$ in DM should have opposite sign. In other words, the group velocity of the pump in the DM should lie between the group velocities of the signal and idler fields.
For the four waves involved in the SFWM, it is straightforward to have the relation $\lambda_{i0}< \lambda_{p0}<\lambda_{s0}$, so this requirement can be fulfilled in a common isotropic DM like silica.

We also note that the two-stage NLI scheme based on spontaneous parametric down conversion (SPDC) in $\chi^{(2)}$ nonlinear media was used to generate photon pairs with symmetric JSF \cite{URen06PRL}. Equation (\ref{DeltakDMexpnd1}) is also valid for analyzing the DM-induced phase shift for the NLI based on SPDC. However, the difference is that in SPDC process the wavelengths of both signal and idler photons are longer than that of the pump. So the group velocities of both signal and idler are usually greater than that of the pump in a common isotropic DM. As a result,  $\tau_s$ and $\tau_i$ have the same sign. In this situation, $0 ^\circ < \rho < 90 ^\circ$ cannot be fulfilled by using an isotropic DM, unless a birefringent DM is employed~\cite{URen05LP}.

We next discuss the small frequency detuning case of $|\omega_{p0}-\omega_{s0(i0)}| \ll \omega_{p0}$. Such kind of SFWM can take place in dispersion-shifted fibers or standard single mode fibers \cite{xylOE, Hall09OE}. In this case, the linear terms cancel out because $k_{DM}^{(1)}(\omega_{p0})\approx k_{DM}^{(1)}(\omega_{s0}) \approx k_{DM}^{(1)}(\omega_{i0})$ and Eq. (\ref{DeltakDMexpnd})
is rewritten as
\begin{equation} \label{DeltakDMexpnd2}
\Delta k_{DM} \approx \frac{k_{DM}^{(2)}(\omega_{p0})}{4} (\Omega_s-\Omega_i)^2,
\end{equation}
and the interference factor becomes
\begin{equation}\label{cos}
{\rm cos}\theta = {\rm cos} \left[\frac{L_{DM} k_{DM}^{(2)}(\omega_{p0})}{8} (\Omega_s-\Omega_i)^2 \right].
\end{equation}
By assuming the DM is a 75-m-long silica fiber and $\lambda_{p0}=1550$ nm, we plot $\left| {\rm cos}\theta \right|^2$.  As shown in Fig. \ref{Figcos}(b), the contour of $\left| {\rm cos}\theta \right|^2$ also has an appearance of parallel stripes, but the orientation angle of the stripes $\rho$ is fixed at 45$^\circ$. Additionally, the width between two adjacent stripes gradually decreases from the center to the two sides due to quadratic dependence upon the frequency difference between the signal and idler photon pairs.
Comparing the two plots in Fig. \ref{Figcos}, one sees that the DM-induced phase shift for the cases of small frequency detuning is different from that of large frequency detuning. In the next subsection, we will focus on studying how to flexibly engineer the JSF by using the interference factor $\left| {\rm cos}\theta \right|^2$ with 45$^\circ$-oriented stripes.

\subsection{Simulation results}

In our simulation, we use single-mode dispersion-shifted fibers (DSF) and a standard single-mode fiber (SMF) as the NF and DM, respectively. The experimental realization of this configuration is straightforward \cite{xylOE,Renyong-OE05,Su19OE}. The wavelengths of the signal, idler, and pump fields are all in the 1550 nm telecom band.
Although our simulations will be performed in the angular frequency space, for the sake of convenient demonstration, the results will be presented in the wavelength space, e.g., the JSF will be plotted as a function of the signal and idler wavelengths, $\lambda_s$ and $\lambda_i$, and the optical bandwidths will be specified in terms of wavelength.

To calculate the JSF, we first deduce the phase shift induced by SMF and the wave vector mismatch in DSF, i.e., $\Delta\phi_{DM}$ and $\Delta k$.
Based on Eq. (\ref{DeltakDMexpnd2}), we can write the phase shift induced by SMF as
\begin{eqnarray}\label{phi-SMF}
\Delta\phi_{DM}=\frac{\lambda_{p0}^2 D_{SMF}L_{DM}}{8 \pi c}(\omega_s-\omega_i )^2,
\end{eqnarray}
where $\lambda_{p0}$ is the central wavelength of pump, and $D_{SMF}$ is the  group velocity dispersion (GVD) coefficient at $\lambda_{p0}$.
Considering the higher-orders dispersion is more significant in DSF, we keep the third and lower order terms of the Taylor series and arrive
\begin{eqnarray}\label{Dk-DSF}
\Delta k = \frac{k^{(2)}_{p0}}{4}( \omega_s - \omega_i )^2+\frac{k^{(3)}_{p0}}{8}(\omega_s+\omega_i-2\omega_{p0})(\omega_s - \omega_i )^2-2\gamma P_p,
\end{eqnarray}
with $k^{(2)}_{p0}=\frac{\lambda_{p0}^2}{2 \pi c}D_{slope}(\lambda_{p0}-\lambda_z)$ and $k^{(3)}_{p0}=-\frac{\lambda_{p0}^4}{(2 \pi c)^2}D_{slope}$, where $\lambda_{z}$ is the zero GVD wavelength of DSF and $D_{slope}$ is the GVD slope at $\lambda_{z}$.
We list below the detailed parameters in the simulation. The pump is Gaussian-shaped with central wavelength $\lambda_{p0}$=1548.5 nm, linear chirp parameter $C_p$=0, and bandwidth (full width at half maximum, FWHM) $\Delta \lambda_p$=1 nm. The DSFs have a zero GVD wavelength $\lambda_{z}$=1548.2 nm with GVD slope $D_{slope}$=0.075 ps/(km$\cdot$nm$^2$), and the nonlinear self phase modulation term $\gamma P_p$=1 km$^{-1}$. The length of each DSF is $L$=50 m. As for the SMF, the GVD coefficient is $D_{SMF}$=17 ps/(km$\cdot$nm) at $\lambda_{p0}$, and the length is $L_{DM}$=7 m.
By substituting the parameters into Eqs. (\ref{phi-SMF}) and (\ref{Dk-DSF}), we can calculate the JSF at the output of the NLI by using Eq. (\ref{NLI-JSF}).

\begin{figure}[htb]
 \includegraphics[width=14.5 cm]{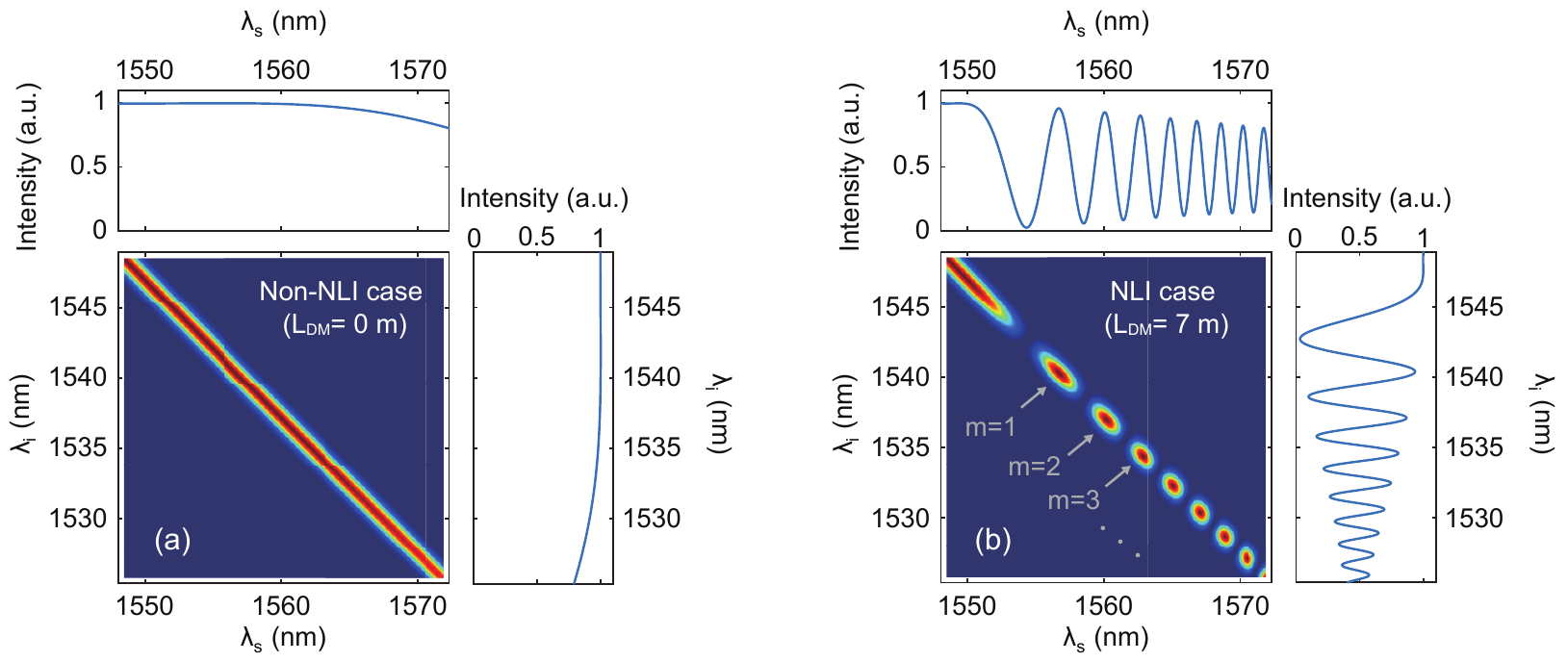}
  \caption{The contour plots of JSF for (a) the non-NLI case of a single-piece 100-m-long DSF ($L_{DM}$=0 m) and (b) the NLI case with 7-m-long DM placed in between two 50-m-long DSFs ($L_{DM}$=7 m). The marginal intensity distributions are plotted next to the corresponding axes.
}\label{p-JSF}
 \end{figure}

For the convenience of comparison, we first perform calculation for the non-NLI case of a single-piece 100-m-long DSF, which is equivalent to the NLI case but with the SMF being removed and the two DSFs being connected directly.
The JSF of the non-NLI case ($L_{DM}$=0 m) is shown in Fig. \ref{p-JSF}(a), exhibiting a strong frequency anti-correlation between the signal and idler bands.


In the NLI case ($L_{DM}$=7 m) shown in Fig. \ref{p-JSF}(b), due to the interference factor $\cos\theta$, the JSF follows a quasi-periodically varying interference profile and exhibits some kind of ``islands" pattern.
The maxima of the islands correspond to the maximum-amplitude points of $\cos\theta$ with $\theta=m\pi$ ($m=0, 1, 2, ...$), while the valleys correspond to zero-amplitude points at $\theta=\pi/2\pm m\pi$ ($m=0, 1, 2, ...$).
The central wavelengths and widths (along the symmetric line with orientation of about $-45 ^\circ $) of each island are mainly determined by the DM-induced phase shift $\Delta \phi_{DM}$.
The quasi-periodicity is because $\Delta \phi_{DM}$ quadratically depends on the frequency detuning between signal and idler photon pairs (see Eq. (\ref{phi-SMF})).
For convenience of discussion,  we label the islands of the JSF with number $``m"$, starting from the zero-detuning point of $m$=0 and labelling the first whole island as $m$=1, and so on, as shown in Fig. \ref{p-JSF}(b). We also denote the central wavelength of $m$th island in the signal (idler) band by $\lambda^{(m)}_{s0(i0)}$. This numbering rule will be adapted in the rest of this paper. For the $m$=1 island in Fig. \ref{p-JSF}(b), we find $\lambda^{(1)}_{s0(i0)}$=1556.7 (1540.4) nm.

We then examine the interference pattern of individual signal and idler bands by calculating the marginal spectral distribution, $F_{s(i)}(\omega_{s(i)})$, which is the projection of JSF on the signal (idler) axis: $F_{s(i)}(\omega_{s(i)})=\int  d \omega_{i(s)} F(\omega_s,\omega_i)$. The results are shown by the curves next to the corresponding axes of JSF.
From Fig. \ref{p-JSF}(b), one sees the quasi-periodically varying interference profile in the marginal distribution. This type of interference in frequency domain was observed before in phase-sensitive fiber amplifier \cite{Renyong-OE05} and inhomogeneous fibers \cite{cui2012PRA}.
It is worth noting that unlike the case of using a single frequency continuous-wave laser as the pump \cite{Renyong-OE05}, the fringe patterns presented here in the signal and idler bands for the pulse-pumped NLI are asymmetrical. This asymmetry originates from the higher order dispersion of the NF (see Eq. (\ref{Dk-DSF})). Moreover, the visibility of the fringe decreases with the increase of the island number $m$. The non-zero values at the minimum is due to the spectral overlap of adjacent islands. More overlap occurs as $m$ increases. The overlap can be seen as a consequence of pulsed pumping.

The results in Fig. \ref{p-JSF} clearly demonstrate that the JSF from NLI is modified by the phase shift induced by the DM (i.e., the SMF).
With the increase of $m$,
the frequency correlation of each island is changed from negatively-correlated, to un-correlated, and eventually to positively-correlated. The reason behind the phenomenon is that
the periodicity for the contour of the interference factor $\left| {\rm cos}\theta \right|^2$ decrease with the increase of detuning (see Fig. \ref{Figcos}(b)), or in other word, the width between two adjacent stripes of $\left| {\rm cos}\theta \right|^2$ decreases with the increase of number $m$.
Because the signal and idler photon pairs are amplified or de-amplified in pairs, this interference patten of NLI can be viewed as a multi-channel band-pass filtering. Different from the passive filtering used in reshaping the JSF of photon pairs (as discussed in Eq.(\ref{TPSf})), the filtering effect in NLI is active and will not introduce loss and un-correlated noise photons. As we will see later, this type of active filter can improve the mode purity of the photon pairs but without the reduction in the collection efficiencies.


\subsection{Mode structure}

With the modified JSF in Eq.(\ref{NLI-JSF}), let us examine the mode structure for the fields from the NLI and compare it to that without the DM. First of all, since the JSF from NLI is divided into islands, which are separated from or orthogonal to each other, each island can be viewed as an individual JSF by filtering so that the state in Eq.(\ref{TPS}) can be rewritten as
\begin{eqnarray}\label{TPS-Sm}
|\Psi\rangle \approx |vac\rangle + G_m\sum_{m=0} |\Psi_2^{(m)}\rangle,
\end{eqnarray}
where $G_m = G{\cal N}_m$ and
\begin{eqnarray}\label{TPS-m}
|\Psi_2^{(m)}\rangle = {\cal N}_m^{-1}\int d \omega_s d\omega_iF(\omega_s,\omega_i)f_s^{(m)}(\omega_s) f_i^{(m)}(\omega_i) \hat a_s ^{\dag}(\omega_s)\hat a_i^{\dag}(\omega_i)|vac\rangle
\end{eqnarray}
is the two-photon state of the $m$th island with
\begin{eqnarray}\label{N-m}
 {\cal N}_m^{2}\equiv \int d \omega_s d\omega_i|F(\omega_s,\omega_i)|^2|f_s^{(m)}(\omega_s) f_i^{(m)}(\omega_i)|^2
\end{eqnarray}
as the normalization factor and $f_s^{(m)}(\omega_s)$, $f_i^{(m)}(\omega_i)$ as the proper filter functions to isolate the $m$th island in JSF. More specifically, we use the rectangular-shaped filter functions
\begin{equation}\label{filter}
f^{(m)}_{s(i)}(\omega_{s(i)})=
\left\{
\begin{array}{ll}
  1, \mathrm{if} \left|\omega_{s(i)}-\omega_{s0(i0)}^{(m)}\right| \leq \frac{\sigma_{s(i)}}{2} \\
  0, \mathrm{if} \left|\omega_{s(i)}-\omega_{s0(i0)}^{(m)}\right| > \frac{\sigma_{s(i)}}{2}
\end{array}
\right.
\end{equation}
with $\omega_{s0(i0)}^{(m)}$ as the central frequency of the $m$th island and $\sigma_{s(i)}$ as the bandwidth of the filter.

The two-photon state in Eq.(\ref{TPS-Sm}) is an entangled state of multiple frequency components \cite{lu03} and can be viewed as in the form of multi-dimensional entangled states with each island representing a component in the high dimensional space \cite{Morandotti17}. However, this view relies on that each island in the JSF represents a single-mode two-photon state, which is exactly what we would like to achieve with our NLI. To find the mode property of each island, we examine next the modal purity of each island in the JSF from the NLI.

\subsection{Modal purity and collection efficiencies}

To examine the modal purity of the islands in JSF of the NLI, we now calculate the intensity correlation function $\bar g_{s(i)}^{(2)}$ for the individual signal (idler) photons of an island, which is isolated out by filters. Since the one-side-filtered $\bar g_{s(i)}^{(2)}$ sets a lower bound for the two-side filtered $\bar g^{(2)}$, which is directly related to the Schmidt number $K$ and describes the modal purity of the filtered photon pairs (see Sect. IIC), we calculate $\bar g_{s(i)}^{(2)}$ for the first three islands in Fig.\ref{p-JSF}(b) by using Eq.(\ref{g2-f}) with the JSF and rectangular filter function described by Eqs.(\ref{NLI-JSF}) and (\ref{filter}), respectively. Both filters in the signal and idler fields are assumed to have the same bandwidth, i.e., $\sigma_{s}=\sigma_{i}=\sigma_{f}$, where $\sigma_{f}$ denoting the common filter bandwidth. As a comparison, we also calculate $\bar g_{s(i)}^{(2)}$ for the non-NLI case. In the non-NLI case, there is no island structure and the marginal distributions are relatively flat within the plotted range, therefore, without loss of generality, we use the same filters as that for the $m$=1 island.

\begin{figure}[htb]
 \includegraphics[width=12.5cm]{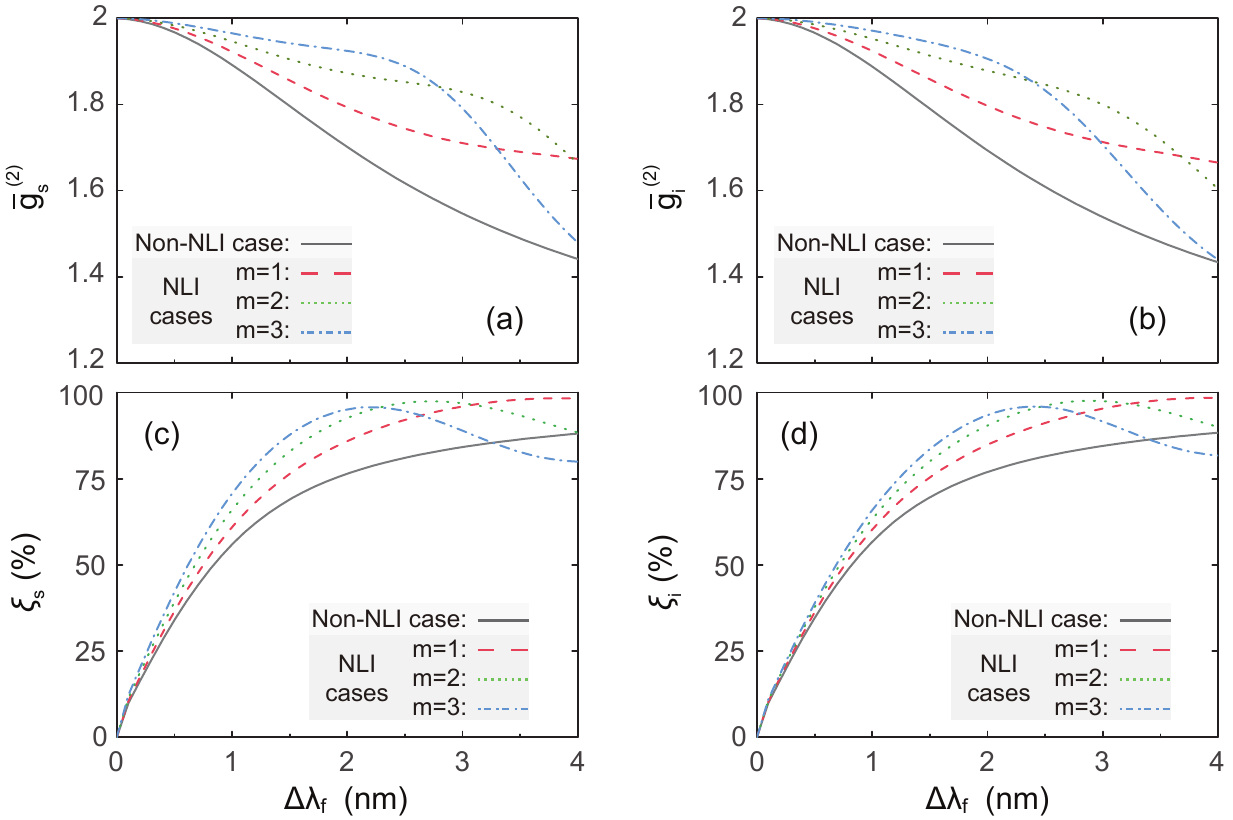}
 \caption{Calculated one-side-filtered second-order intensity correlation functions and collection efficiencies as functions of the common filter bandwidth $\Delta\lambda_{f}$. (a) and (b) show the one-side-filtered $\bar g_{s}^{(2)}$ and $\bar g_{i}^{(2)}$, (c) and (d) show the collection efficiencies $\xi_s$ and $\xi_i$. The dashed, dotted, and dash-dotted curves are the results for three NLI cases with island number $m$=1, 2, and 3, respectively, while the solid curves are results for the non-NLI case.
 }
 \label{g2f}
 \end{figure}

Figures \ref{g2f}(a) and \ref{g2f}(b) respectively present the calculated $\bar g_{s}^{(2)}$ and $\bar g_{i}^{(2)}$ as functions of the common filter bandwidth in terms of wavelength, $\Delta\lambda_{f}$. Here we have used the approximate relation $\Delta\lambda_{f}=\frac{(1550~\mathrm{nm})^2}{2 \pi c}\sigma_{f}$ for the 1550 nm band filters. The dashed, dotted, and dash-dotted curves are the result for islands with island numbers $m$=1, 2, and 3, respectively, while the solid curves are the result for the non-NLI case.
Comparing the results of the signal and idler fields, we find their general trends are similar, except the differences originated from the spectral asymmetry depicted in Fig.\ref{p-JSF}(b).
Therefore, in the following discussion we will focus only on the results of the signal field.

One sees from Fig. \ref{g2f}(a) that $\bar g_{s}^{(2)}$ of all the four cases are very close to 2 when $\Delta\lambda_{f}<$0.2 nm, showing the powerful mode-cleaning effect of an extremely narrow bandpass filter.
With the increase of $\Delta\lambda_{f}$, the advantage of NLI becomes significant.
In some certain range of $\Delta\lambda_{f}$, $\bar g_{s}^{(2)}$ of the NLI cases are higher than that of the non-NLI case, which means an improvement of the modal purity.
Moreover, for each case in Fig. \ref{g2f}(a), $\bar g_{s}^{(2)}$ decreases with the increase of $\Delta\lambda_f$, but the descent rate for each case is different.
Particularly, one sees that there exists a plateau before the sharp drop of $\bar g_{s}^{(2)}$ in the NLI cases whereas in the non-NLI case $\bar g_{s}^{(2)}$ decreases with a nearly constant rate. The plateaus can be seen as the results of the island structure of the interference pattern, while the sharp drop after each plateau is because the components of adjacent islands are also collected as the filter bandwidth increases.
The turning point of the sharp drop of $\Delta\lambda_{f}$ is determined by the valley-to-valley width of the specific island. For example, the turning point for the $m$=2 island is approximately at $\Delta\lambda_{f}$=3 nm.

As discussed in Sect.IIC, collection efficiency $\xi_{s(i)}$ is another important factor in obtaining high purity heralded single-photon source. It has an opposite trend to $\bar g_{s(i)}^{(2)}$ as the bandwidth of the filters changes. Now let us examine how the collection efficiencies are affected in the selection of a specific island in the JSF of NLI by filtering.
Equation (\ref{collection}) is used for the calculation of $\xi_{s(i)}$ with $\eta_{s(i)}=1$.
Again, we perform the calculation for the three NLI cases (the islands with numbers $m$=1, 2, and 3) as well as the non-NLI case. Figures \ref{g2f}(c) and \ref{g2f}(d) respectively show the calculated $\xi_s$ and $\xi_i$ as functions of the common filter bandwidth $\Delta\lambda_{f}$. From the results of the signal field shown in Fig. \ref{g2f}(c), one sees that although the general trends of $\xi_{s}$ are still opposite to those of $\bar g_{s}^{(2)}$ due to the detrimental effect of passive filtering, $\xi_s$ for the NLI cases are in general significantly higher than that of the non-NLI case.
In particular, there exists a maximum value of $\xi_{s}$ for the NLI cases, corresponding to the plateau turning point of $\bar g_{s}^{(2)}$ in Fig. \ref{g2f}(a).
The existence of the maximum of $\xi_{s}$ is due to the same reason for the tuning point of $\bar g_{s}^{(2)}$ that some uncorrelated photons from the adjacent islands are also collected by the filter.
The maxima of $\xi_{s}$ are about 98\%, 97\%, and 96\% for the islands with $m$=1, 2, and 3, respectively. Notice this decreasing trend of the maxima of $\xi_{s(i)}$ with $m$ is in accordance with the drop of fringe visibility in Fig. \ref{p-JSF}(b).
Therefore, we believe the improvement of $\xi_{s}$ by using NLI is originated from the active filtering effect. Higher visibility of interference fringe means less uncorrelated photons being collected by the filter.

Finally, inspecting Figs.\ref{g2f} (a)-(d) together, we find there is an optimum value for the filter bandwidth $\Delta\lambda_{f}$, with which the specific island in JSF can be properly isolated so that both $\bar g_{s(i)}^{(2)}$ and $\xi_{s(i)}$ are relatively high. For example, photon pairs extracted from the $m$=3 island gives the best number of $\xi_{s}$ = 96\% with $\bar g_{s}^{(2)}$=1.91. Moreover, we recently implemented the two-stage NLI experiment and demonstrated the good agreement between the experimental results and simulations \cite{Su19OE}. Therefore, applying optical filter with proper bandwidth at the output of NLI does not harm the collection efficiency as much as the non-NLI case. The less-than-ideal performance is because the different islands in the JSF of Fig.\ref{p-JSF}(b) do not separate far enough to have a clean cut for the filters. This leaves us rooms for further improvement, as will be analyzed in Sect. IV.

\subsection{Condition for obtaining an isolated island with factorable spectrum}

In the small-detuning case discussed above, one sees the exact shape of the islands in the interference-modified JSF is a manifestation of the pump profile (with a $-45^{\circ}$ oriented stripe pattern) and interference term (with a $45^{\circ}$ oriented stripe pattern, as shown in Fig.\ref{Figcos}(b)) while the sinc-function plays a less role because the phase matching condition can be approximated as ${\rm sinc}\big(\frac{\Delta k L}{2}\big)\approx1$. In this case, a factorable island should be round-shaped.
Let's analyze how to realize the JSF having a round island. Fig. \ref{p-JSF}(b) shows that
the main maxima occur at $\theta = m\pi$ with $m=1,2,...$ as the order number. For the interference term in Eq.(\ref{cos}), by defining small variations $\Delta\Omega_{s(i)} \equiv \Omega_{s(i)} - \Omega_{s(i)}^{(m)}$ around the maxima $\Omega_{s(i)}^{(m)}$ which satisfy $L_{DM} k_{DM}^{(2)}(\omega_{p0}) (\Omega_s^{(m)}-\Omega_i^{(m)})^2/8 = \theta = m\pi$, we can approximate $\cos\theta$ around the maxima by a Gaussian: $\exp[-(\Delta\Omega_{s}-\Delta\Omega_{i})^2/(2\sigma_{int}^2)]$, where $\sigma_{int}^2$ is the approximate width of the $45^{\circ}$ oriented stripe pattern (see Fig.\ref{Figcos}(b)). To find $\sigma_{int}^2$, we make a Taylor expansion of function $\left|\cos ax^2\right|$ around maxima $ax_m^2=m\pi (x=x_m+\Delta x)$:
\begin{eqnarray}\label{cos-tay}
\cos ax^2\approx 1 - 2 m\pi a (\Delta x)^2 \approx 1- (\Delta x)^2/(2\sigma_{int}^2) \approx e^{- (\Delta x)^2/(2\sigma_{int}^2)},
\end{eqnarray}
which gives $\sigma_{int}^2 \approx 1/(4m\pi a)$, indicating that the stripes get narrower as $m$ increases, as illustrated in Fig.\ref{Figcos}(b). Using Eq.(\ref{cos}), we find $a=L_{DM} k_{DM}^{(2)}(\omega_{p0})/8$ so that $\sigma_{int}^2 \approx 2/[m\pi L_{DM} k_{DM}^{(2)}(\omega_{p0})]$. In order to have a round shape for the $m$-th island, the $-45^{\circ}$ oriented stripe pattern must have the same width as the $45^{\circ}$ oriented stripe pattern. This leads to the following condition
\begin{eqnarray}\label{pump-length}
2\sigma_p^2 = \sigma_{int}^2 = 2/[m\pi L_{DM} k_{DM}^{(2)}(\omega_{p0})]
\end{eqnarray}
or
\begin{eqnarray}\label{pump-length2}
\sigma_p^2 L_{DM}  = 1/[m\pi  k_{DM}^{(2)}(\omega_{p0})]
\end{eqnarray}
for a round $m$-th island of the modified JSF. Therefore, the shape of each island can be tuned by properly adjusting the pump bandwidth ($\sigma_p$) and the length of the dispersive medium ($L_{DM}$).

On the other hand, an island with a factorable spectrum may not necessarily be round-shaped. In general, a factorable island can have an elliptic shape with the major and minor axes respectively parallel to the two axes of JSF \cite{Grice-PRA-2001}.
For example, in the large-detuning case depicted in Fig. \ref{Figcos}(a), orientation for the strips of the interference factor $|{\rm cos}\theta|^2$ is not perpendicular to that of the pump.  From Fig. \ref{Figcos}(a), one sees that $|{\rm cos}\theta|^2$ has a uniform periodic pattern, and the maxima occurs at $\theta=n\pi$ ($n$ is an arbitrary integer).
We can approximate the interference factor in Eq. (\ref{eqcoslar}) around one of its maxima $\Omega_{s(i)}^{(n)}$ by a Gaussian:
\begin{equation}
{\rm cos}\theta \approx {\exp} \left[-\frac{(\tau_s \Delta\Omega_s + \tau_i \Delta\Omega_i)^2}{8 L_{DM}^{-2}} \right],
\end{equation}
where $\Delta\Omega_{s(i)}=\Omega_{s(i)}-\Omega_{s(i)}^{(n)}$.
Using the pump envelop function in Eq. (\ref{JSF-fb-s}), the spectrum of the island can be expressed by the product of the pump envelop and the interference factor:
\begin{eqnarray}\label{eqJSFellip}
F(\Delta\Omega_s,\Delta \Omega_i) \propto \exp\big[-\frac{(\Delta\Omega_s+\Delta\Omega_i)^2}{4\sigma_p^2}-\frac{(\tau_s \Delta\Omega_s + \tau_i \Delta\Omega_i)^2}{8 L_{DM}^{-2}}\big].
\end{eqnarray}
To eliminate the spectral correlation, the term $\Delta\Omega_s \Delta\Omega_i$ in Eq. (\ref{eqJSFellip}) should vanish. This leads to the condition for obtaining a factorable island:
\begin{equation}\label{eqellip}
\sigma_p^2=-2/(\tau_s \tau_i L_{DM}^2),
\end{equation}
By substituting Eq. (\ref{eqellip}) into Eq. (\ref{eqJSFellip}), we arrive at the factorable spectra
\begin{eqnarray}\label{eqJSFellip2}
F(\Delta\Omega_s,\Delta\Omega_i) \propto \exp\big[-\frac{\tau_s \Delta\Omega_s^2 - \tau_i \Delta\Omega_i^2}{8 L_{DM}^{-2}(\tau_s-\tau_i)^{-1}}\big],
\end{eqnarray}
corresponding to an elliptically shaped island with the ellipticity determined by the ratio $|\tau_s/\tau_i|$. 
It is straightforward to show that for the special case of $\tau_s=-\tau_i$ ($\rho=45 ^\circ$), i. e., the contour of the interference factor is perpendicular to that of the pump, Eq. (\ref{eqellip}) has the simplified form: $2/(\tau_s L_{DM})^2= \sigma_p^2$. In this case, the strip width of the interference factor matches that of the pump envelope and the island is round-shaped.

\begin{figure}[htbp]
	\includegraphics[width=14.5cm]{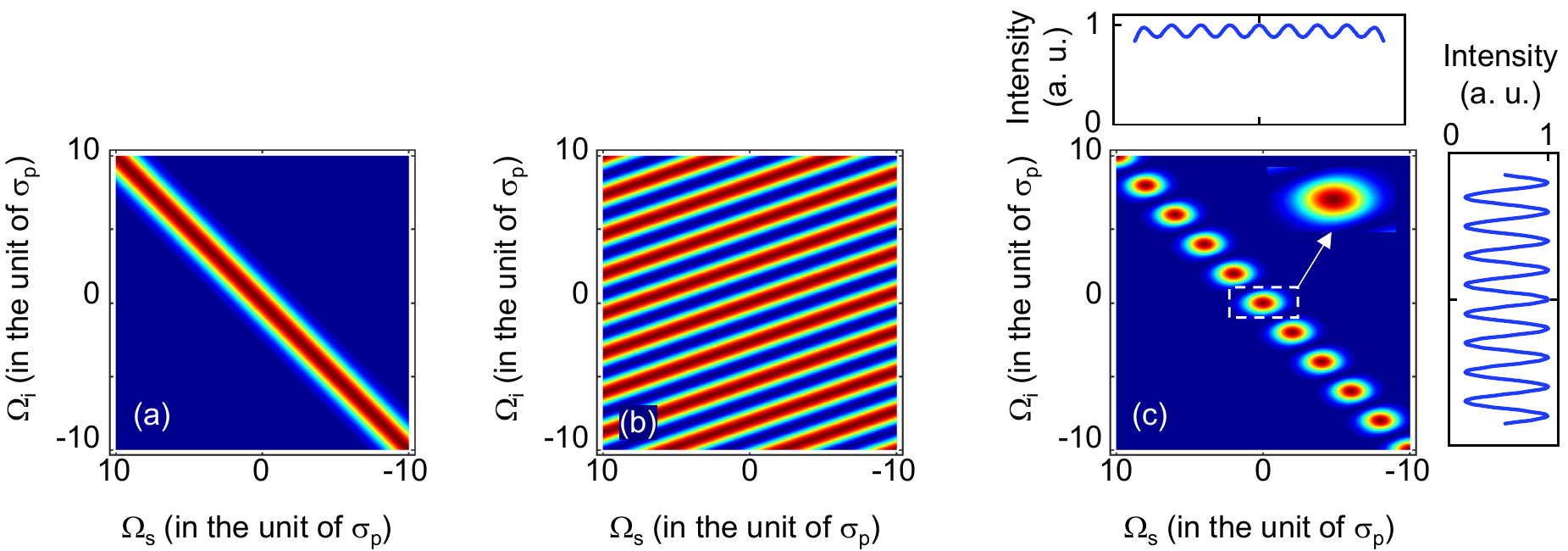}
	\caption{Contour plots of (a) pump envelope, (b) interference factor $|{\rm cos}\theta|^2$, and (c) corresponding JSF of NLI. The marginal intensity distributions in individual signal and idler fields are plotted next to the corresponding axes of the JSF.
	}
	\label{FacJSF}
\end{figure}

Figure \ref{FacJSF} (c) shows a typical JSF formed by a series of elliptical islands, from which spectral factorable photon pairs can be obtained. In the simulation, the pump bandwidth and the length of the DM (i. e., the silica fiber) are properly set to satisfy the condition in Eq. (\ref{eqellip}). Figs. \ref{FacJSF}(a) and \ref{FacJSF}(b) respectively show the contours of the pump envelop and interference factor of NLI.
From Fig. \ref{FacJSF}(c), one sees that the orientation of the major or minor axis of each elliptically shaped island is parallel to one axis of JSF. To show the active filtering effect of the NLI, we also plot the interference patterns of individual signal and idler fields by calculating their marginal spectral distribution, as shown by the curves next to the corresponding axes of JSF. Note that in Fig. \ref{FacJSF}(c), the visibility of the interference in the signal field is obviously lower than that in the idler field because there is a large overlap between adjacent islands in the projection of the signal field.
It thus is hard to isolate and select out the elliptically factorable island with filters. In this case, the best we can do is to make the island into round shape by making the orientation of the strip 45 degree. On the other hand, the elliptical shape will work if we can make islands well separated so that there is no overlap between adjacent islands when we select them out with filters. This is the topic of discussion in the next section.

\section{FURTHER ENGINEERING FOR BETTER CONTROL OF JSF}

In the above theoretical analysis, we have demonstrated that the JSF of photon pairs can be engineered to have some sort of island pattern by introducing a phase shift in two-stage NLI with a DM.
We also find that the overlapping of the adjacent islands is detrimental to creating JSF from which a spectrally factorable island can be isolated out.
In this section, we will respectively resort to two different methods, namely, the programmable optical filtering technology and the multi-stage NLI scheme, to realize more flexible and precise engineering of JSF. Using these methods, we can create JSF with island patterns that are more factorable and sufficiently-isolated, which is desirable in generating multi-dimensional entanglement. Moreover, we will also discuss how to make full use of each island of the JSF to achieve multi-channel outputs.

\subsection{Using programmable optical filter for arbitrary spectral engineering}

\begin{figure}[htb]
 \includegraphics[width=15cm]{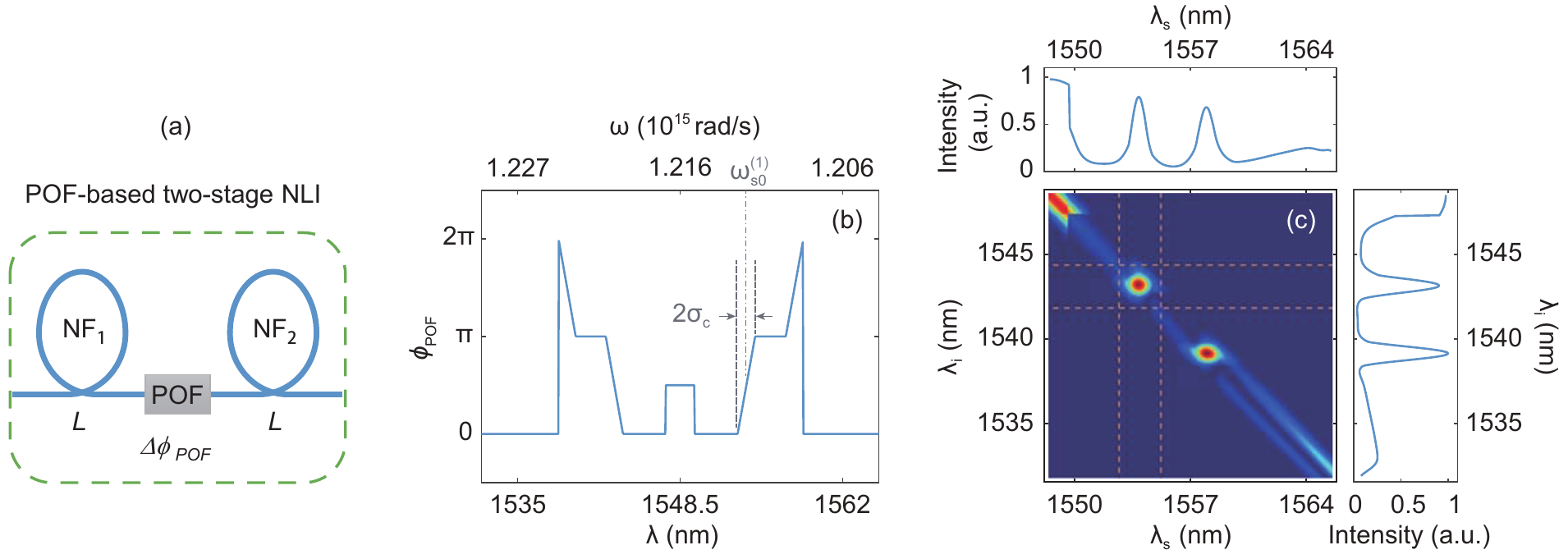}
 \caption{(a) Schematic of POF-based two-stage NLI. (b) Phase function of POF. As an example, the dashed lines mark the phase-control range with width $\sigma_c$ around $\omega^{(1)}_{s0}$, which is the signal central frequency of the $m$=1 island. (c) Contour plot of JSF output from the POF-based NLI.}
\label{POF}
\end{figure}

A programmable optical filter (POF) can introduce arbitrary phase at different frequency (wavelength), which can be described by phase function $\phi_{POF}(\omega)$. If we replace the DM with a POF in the two-stage NLI, as shown in Fig. \ref{POF}(a), the DM-induced phase shift $\Delta\phi_{DM}$ in Eq. (\ref{phase-NF2}) will be accordingly replaced with the POF-induced phase shift $\Delta\phi_{POF}$, then the interference factor in Eq. (\ref{phase-NF2}) becomes
\begin{equation}\label{theta-POF}
\cos\theta=\cos\bigg(\frac{\Delta k L}{2}+\frac{\Delta\phi_{POF}}{2}\bigg),
\end{equation}
with
\begin{equation}\label{DeltaPhi-POF}
\Delta\phi_{POF}=2\phi_{POF}(\omega_p)-\phi_{POF}(\omega_s)-\phi_{POF}(\omega_i). \end{equation}
In this case, we can tailor the JSF with much more flexibility by arbitrarily controlling the phase function $\phi_{POF}(\omega)$ of POF.

Different from the DM-induced phase function $\phi_{DM}(\omega)$, which is a continuous function, we can define $\phi_{POF}(\omega)$ as a piecewise function to increase the flexibility of spectral control.
For example, when $\phi_{POF}(\omega)$ is designed as the function shown in Fig.\ref{POF}(b), the JSF of the output from this POF-based NLI will show two factorable islands, as shown in Fig.\ref{POF}(c). But the design of POF for engineering special JSF is quite complicated. So, we will devote the detailed discussion to another paper \cite{Cui2020}.

We can show in that paper that JSF with sufficiently-isolated and factorable islands at arbitrarily chosen wavelengths can be realized using a POF-based NLI. It paves the way for developing a multi-channel source of photon pairs with high modal purity, high collection efficiency, and arbitrary output wavelengths. However, one problem with POF-based NLI is that the currently available POFs have a relatively high insertion loss, which could limit the performance of the NLI. So, next we will discuss a totally different approach in fine engineering of the JSF.

\subsection{Multi-stage NLI for more precise engineering}

\begin{figure}[htb]
 \includegraphics[width=12.5cm]{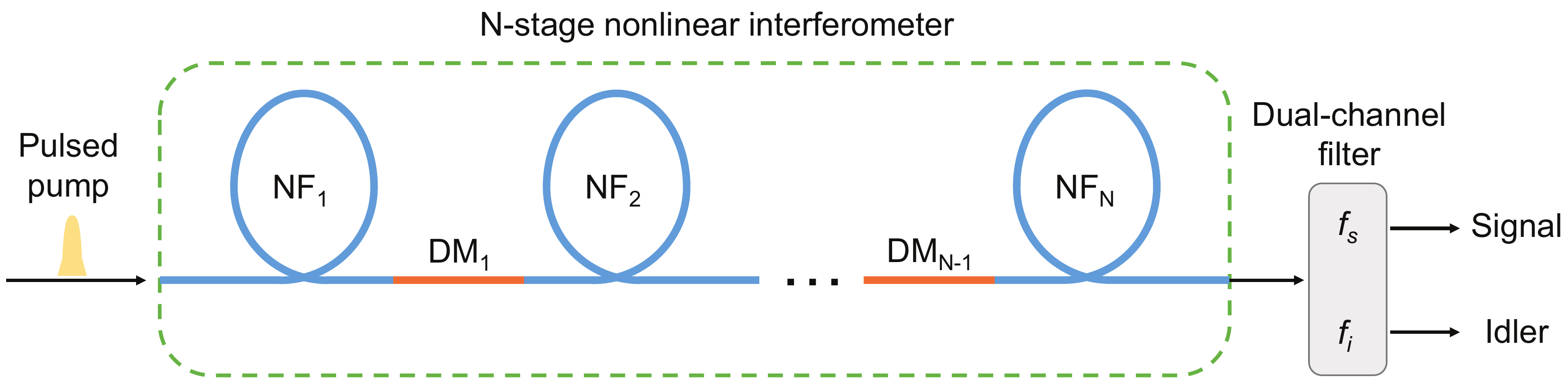}%
 \caption{Schematic of an $N$-stage nonlinear interferometer (NLI) with $N$ pieces of nonlinear fibers (NFs) and $N-1$ pieces of dispersive media (DMs).}
\label{multi}
 \end{figure}

Another method of fine engineering the JSF is to use a multi-stage NLI. As shown in Fig. \ref{multi}, an N-stage NLI consists of $N$ pieces of NFs and $N-1$ pieces of DM placed in between every two NFs. Assume that all the NFs of the multi-stage NLI are identical, and so do DMs. Moreover,  the insertion and transmission losses of NFs and DMs can be neglected. The two-photon state from the NLI with stage number $N$ is then in the form of Eq.(\ref{TPS2}) but with the JSF being modified as
\begin{equation}\label{multi-JSF}
F^{(N)}_{NLI}({\omega _s},{\omega _i}) ={\cal N} \exp\left[ -\frac{(\omega_s+\omega_i - 2 \omega_{p0})^2}{4\sigma^2_p} (1+j C_p)\right] \times \mathrm{sinc} \left(\frac{\Delta kL}{2}\right) \times H(\theta),
\end{equation}
with
\begin{equation}
\label{multi-H}
H(\theta) = \sum^{N}_{n=1} e^{2 j (n-1)\theta}=\frac{\sin N \theta}{\sin \theta} e^{j(N-1)\theta},
\end{equation}
where $H(\theta)$ with $\theta=\frac{\Delta k L}{2}+\frac{\Delta\phi_{DM}}{2}$ (Eq.(\ref{phase-NF2})) is a modulation function similar to the interference factor of a multi-slit interferometer in classical optics.

\begin{figure}[htb]
 \includegraphics[width=15.5cm]{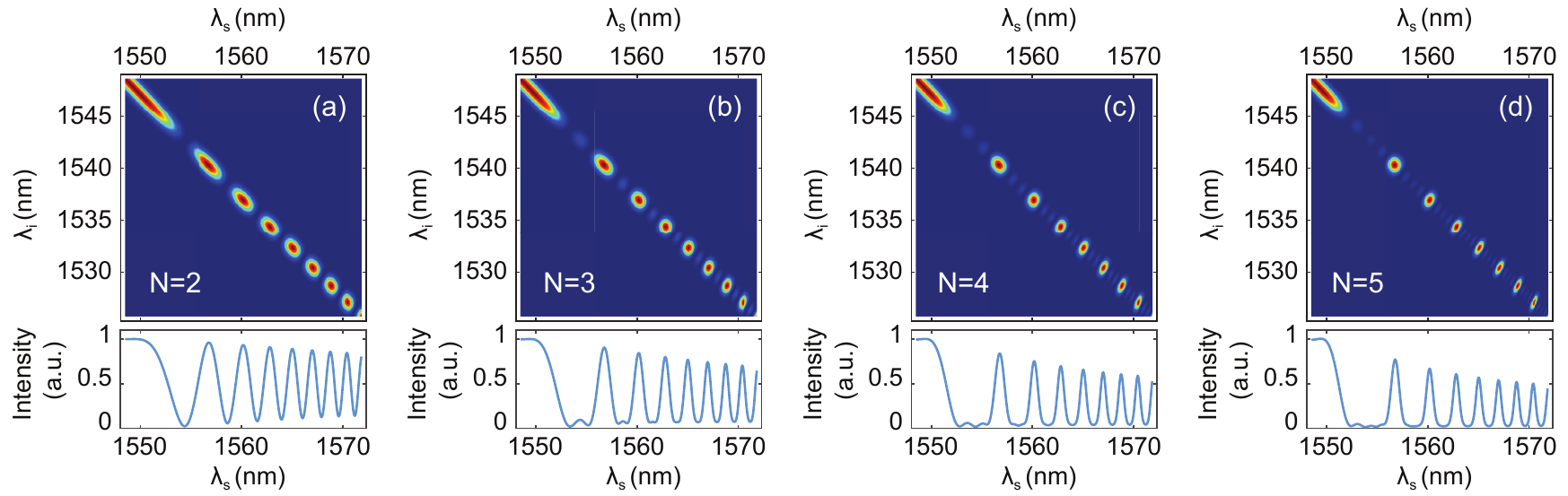}
 \caption{Contour plots of JSF and intensity distributions in the signal band. (a)-(d) are the results of NLIs with stage number $N$=2, 3, 4, and 5, respectively.}
\label{JSF-multi}
 \end{figure}

To demonstrate the performance of the multi-stage NLI, we perform simulations in the same way as that for the two-stage NLI in Sect.III. Again we employ the DSFs (each of length 50 m) and SMFs (each of length 7 m) as the NFs and DMs, respectively. The parameters given in Sect. IIIB are used as well.
Using Eq.(\ref{multi-JSF}), we plot the JSFs of the two-stage($N$=2) and the multi-stage ($N$=3, 4, and 5) NLIs, as shown in Figs. \ref{JSF-multi}(a)-(d), respectively.
From Fig. \ref{JSF-multi}, we find the JSF has the following features.
First, with the increase of stage number $N$, the central wavelengths of the primary islands do not vary but the width of each island (along the symmetric line oriented at about $-45^{\circ}$) decreases. In the four plots,
the central wavelengths of the islands with the same label number are the same: the $m$=1, 2 and 3 islands in the signal (idler) band are centering at 1556.7 nm (1540.4 nm), 1560.2 nm (1537.0 nm), and 1562.8 nm (1534.5 nm), respectively.
Second, in the cases of $N\geq3$, there exists $N-2$ secondary islands between two adjacent primary islands, and the ratio between the intensities of the primary and secondary islands increases with the increase of $N$. As a result, the fringe visibility of marginal intensity distributions in the signal band accordingly increases with $N$ as well. These results indicate that JSF with sufficiently-isolated islands can be realized by a multi-stage NLI. We have recently verified the simulation results by carrying out a three-stage NLI experiment \cite{LJM20arx}.

\begin{figure}[htb]
 \includegraphics[width=12.5cm]{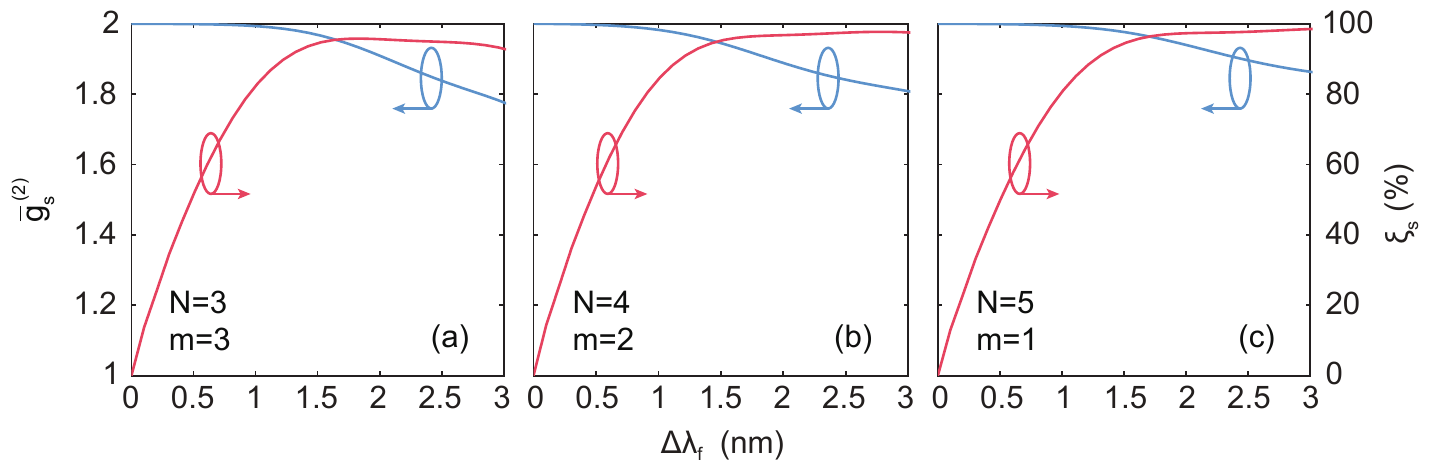}
 \caption{Calculated one-side-filtered intensity correlation function $\bar g_{s}^{(2)}$ and collection efficiency $\xi_{s}$ as function of the common filter bandwidth $\Delta\lambda_f$ for the mostly factorable islands in Figs. \ref{JSF-multi}(b)-(d). The island and stage numbers for the three cases are: (a) $m=3$ and $N=3$, (b) $m=2$ and $N=4$, and (c) $m=1$ and $N=5$.}
 \label{JSF-multi-g2}
\end{figure}

Note that for the DM used in plotting Fig. \ref{JSF-multi}, the feature of interference term is similar to that shown in Fig. \ref{Figcos}(b): with a $45^{\circ}$ oriented stripe pattern and the stripe width decrease with the increase of detuning. In this case, the width of the island in JSF is related to both the labelled number $m$ and stage number $N$.
For an $N$-stage NLI, it is possible to find a round shaped island corresponding to a specific number $m$. With the change of stage number $N$, the label number $m$ of the round shaped island varies.
For example, from Figs. \ref{JSF-multi}(b)-(d), we find the most factorable islands are: (i) $m=3$ island for $N=3$, (ii) $m=2$ island for $N=4$, and (iii) $m=1$ island for $N=5$. To characterize the three cases, we respectively calculate the one-side-filtered intensity correlation function $\bar g_{s}^{(2)}$ and collection efficiency $\xi_{s}$ as functions of the common filter bandwidth $\Delta\lambda_f$ for each island. As shown in Fig. \ref{JSF-multi-g2}, in each case, $\xi_{s}$ rises quickly with $\Delta\lambda_f$ when the bandwidth $\Delta\lambda_f$ is less than 1.5 nm, but approaches to some values close to unity for $\Delta\lambda_f$ in the range of 1.5 nm to 3 nm.
Meanwhile, although $\bar g_{s}^{(2)}$ decreases with the increase of $\Delta\lambda_f$, the descending rate is relatively low and depends on the stage number $N$. We find that $\bar g_{s}^{(2)}>1.95$ and $\xi_{s}>95\%$ can be simultaneously achieved for all the three cases.

\begin{figure}[htb]
 \includegraphics[width=12cm]{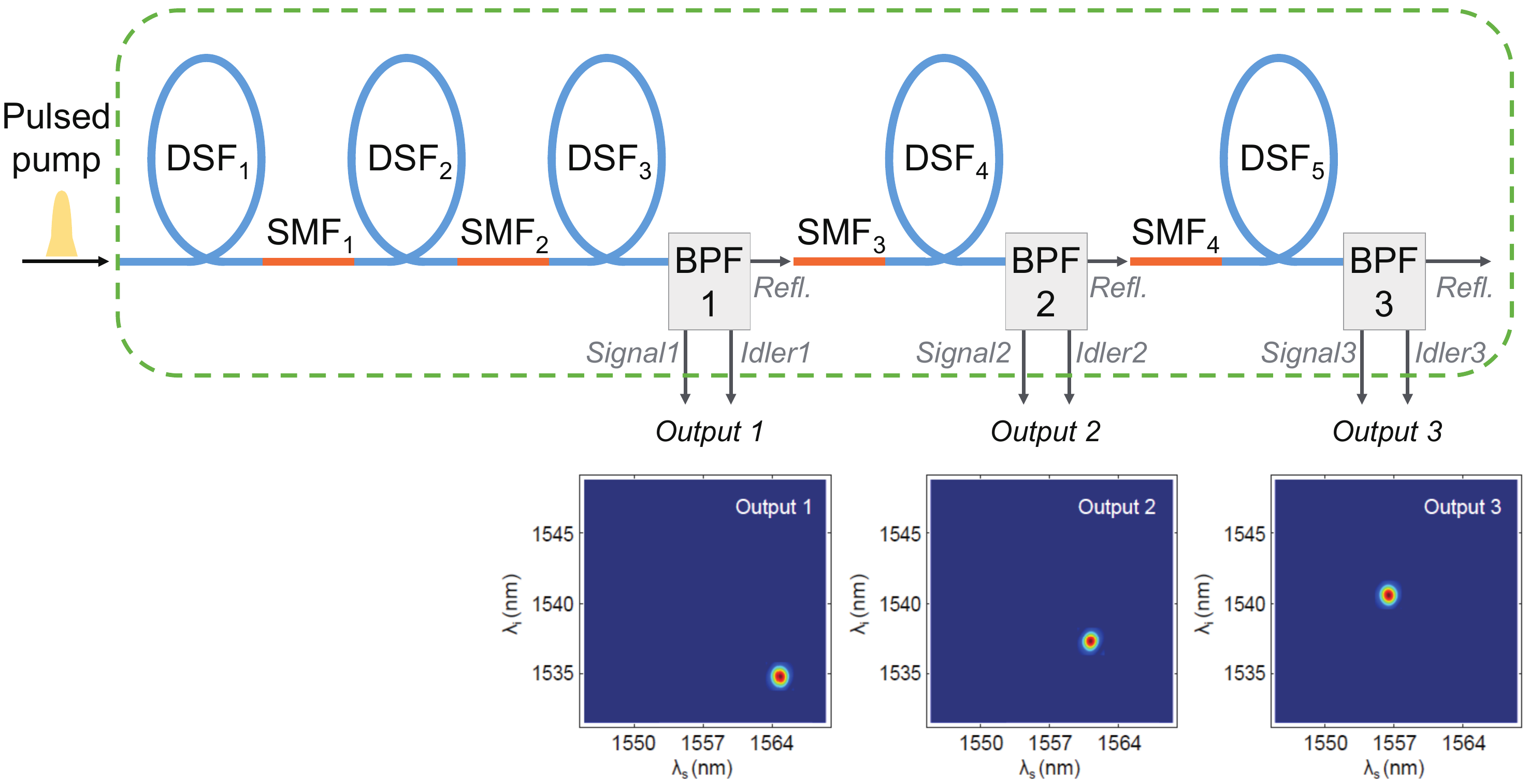}
 \caption{Three-channel source of spectrally factorable photon pairs based on a five-stage NLI. Contour plots of the JSFs for the outputs 1-3 are shown next to the corresponding ports of BPF1-BPF3, respectively. DSF, dispersion shifted fiber; SMF, single-mode fiber, BPF, dual-channel bandpass filter with a reflection port.}\label{multi-out}
\end{figure}

The active filtering effect in NLI is similar to that in an optical parametric oscillator (OPO) far below threshold \cite{Lu00PRA}. However, there is difference between them. OPO is an overall result of the resonant
cavity, but each intermediate stage in the N-stage NLI can be separated. For the $N$-stage NLI in Fig. \ref{JSF-multi},
only one island can be the most factorable for given stage number, but we can make full use of the multiple stages by successively carving out the factorable islands.
Figure \ref{multi-out} depicts a scheme of realizing a multi-channel single photon source with a five-stage NLI, in which three suitable dual-band band-pass filters (BPFs) having both transmission and reflection ports are inserted. The two rectangularly shaped pass bands of BPF1 are centering at 1562.8 nm and 1534.5 nm, respectively, and the bandwidth of each band is 1.5-nm-width.
By placing BPF1 right after DSF3, the $m$=3 island (see Fig. \ref{JSF-multi}(b)) is selected out , and the other fields from the reflected port are sent to the next stage. The contour plot of the JSF for the transmitted port of BPF1 (Output1) is shown in Fig. \ref{multi-out}.
Similarly, the $m$=2 and $m$=1 islands in Fig. \ref{JSF-multi}(c) and (d) are extracted out by respectively placing BPF2 and BPF3 after DSF4 and DSF5. The pass bands of both BPF2 and BPF3 also have rectangularly shaped spectra but with their center wavelengths properly set to fit the center of selected islands. The JSFs of the corresponding outputs (Output2, Output3) are also depicted in Fig. \ref{multi-out}. It can be seen that all the JSFs are nearly round and factorizable. By doing so,
a multi-channel source of photon pair with high purity and efficiency can also be realized, which can be further used to obtain multi-dimensional entanglement \cite{Morandotti17}.

On the other hand, if the feature of interference term induced by a DM is similar to that shown in Fig. \ref{Figcos}(a): the stripe width does not vary with detuning and the orientation angle of strips is in the range of $0 ^\circ < \rho < 90 ^\circ$, so the width of the island is irrelevant to the labelled number $m$ and decreases with the increase of stage number $N$. When the amount of phase shift induced by DM and the stage number of NLI are properly designed, the factorable two-photon state can be efficiently extracted from each island, which is well separated from each other \cite{URen05LP}

It is worth noting that the performance of multi-stage NLI highly depends on its transmission efficiency \cite{Cui20POF}. Although the stage-number of NLI gives another degree of freedom for precisely engineering the spectrum of two-photon state, high stage number may prevent the muti-stage NLI from practical application. Since insertion loss will be inevitably induced by each DM, the transmission efficiency of NLI will decrease with the increase of stage-number $N$. Therefore, if the amount of phase shift induced by each DM is small, say, it is about the same as the phase mismatch $\Delta k L$ in the nonlinear medium, to efficiently obtain spectral factorable two-photon stage, stage number of NLI must be large enough \cite{URen05LP}. If the phase shift of DM is large enough, i.e., $\phi_{DM}\gg \Delta k L$, the goal of achieving an island which is factorable and well separated from other islands can be realized by exploiting the NLI with stage number less than 5.


\subsection{Uneven Multi-stage NLI}

The multi-stage NLI discussed in the previous part still has some mini-maxima in between the main maxima. This contributes to the imperfect $g^{(2)}$ and efficiency. Elimination of these mini maxima is thus expected to improve modal purity and efficiency. Notice that the mini-maxima originate from the H-function in Eq.(\ref{multi-H}), which normally gives (N-1)-th harmonic of $\cos2\theta$, that is, $\cos 2(N-1)\theta$. To change it, we consider uneven length of nonlinear fibers (DSFs) for different part of the multi-stage NLI. To make things simple, we assume the DMs sandwiched in between all have the same length. Furthermore, we can select the fiber parameters for the DSFs so that $\Delta k L/2 \ll 1$ for all and the sinc-function in Eq.(\ref{multi-JSF}) is approximately 1, that is, all the DSFs are nearly phase matched.  Then, after realizing that the contribution from each DSF is proportional to its length when phase matched, we obtain the JSF as
\begin{equation}\label{multi-JSF-un}
F^{(UN)}_{NLI}({\omega _s},{\omega _i}) =\exp\left[ -\frac{(\omega_s+\omega_i - 2 \omega_{p0})^2}{4\sigma^2_p} (1+j C_p)\right] \times K(\theta),
\end{equation}
where the superscript `UN' denotes uneven N-stage NLI and
\begin{equation}
\label{multi-K}
K(\theta) = \sum_{n=1}^N L_n e^{2j(n-1)\theta}
\end{equation}
with $\theta = \Delta k_{DM} L_{DM}/2$ and $L_n (n=1,2,...,N)$ as the length of individual DSF. Similar to the H-function in Eq.(\ref{multi-JSF}), the interference pattern originates from $|K(\theta)|^2$, which normally gives harmonics of $\cos 2\theta$ up to $\cos 2(N-1)\theta$ and $N-1$ mini maxima.

On the other hand, if we can arrange $L_n (n=1,2,...,N)$ in such a way that $|K(\theta)|^2\propto \cos^{2(N-1)} \theta \propto (1+ \cos2\theta)^{N-1}$, it will totally eliminate the mini maxima. This leads to the following equation:
\begin{equation}
(1+e^{2j\theta})^{N-1} = \sum_{n=1}^{N} (L_n/L_1) e^{2j(n-1)\theta},
\end{equation}
or
\begin{eqnarray}
\label{Ln}
L_n= L_1 C_{N-1}^{n-1}=L_1 \frac{(N-1)!}{(n-1)!(N-n)!},
\end{eqnarray}
which is a binomial distribution. So, if the different sections of NLI follow a binomial pattern in length, then $|K(\theta)|^2 \propto (1+\cos 2\theta)^{N-1}$, which has no mini maxima. In the meantime, the width of the main maxima narrow as $N$ increases, leaving different islands in JSF well separated.

With $|K(\theta)|= |\cos^{N-1}\theta|$ and approximation in Eq.(\ref{cos-tay}), we easily obtain the condition
\begin{eqnarray}\label{pump-length3}
\sigma_p^2 L_{DM}  = 1/[m(N-1)\pi  k_{DM}^{(2)}(\omega_{p0})]
\end{eqnarray}
for a round $m$-th island of the JSF in the uneven $N$-stage NLI.

\begin{figure}[htb]
	\includegraphics[width=9cm]{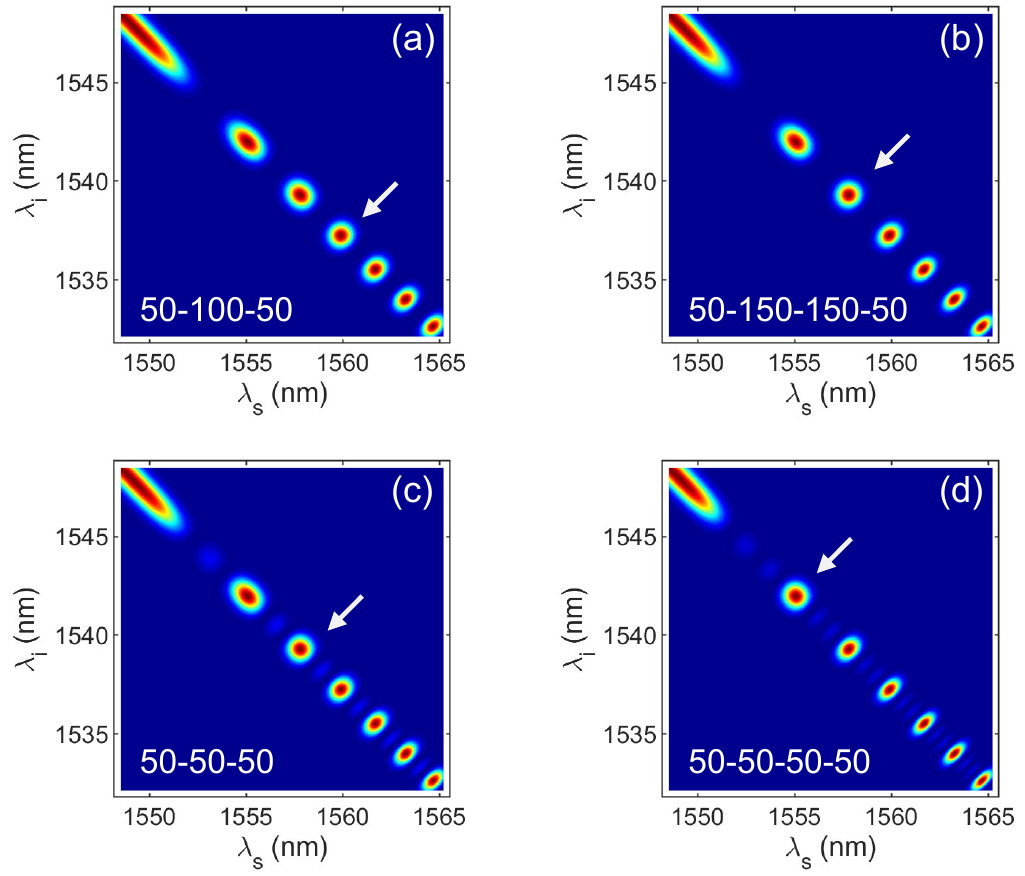}
	\caption{Contour plot for the JSF of (a) 3-stage NLI with uneven 50-100-50 length, (b) 4-stage NLI with uneven 50-150-150-50 length, (c) 3-stage NLI with even 50-50-50 length, and (d) 4-stage NLI with even 50-50-50-50 length. The numbers denote the lengths of individual DSFs in the units of meter. The length of individual SMFs in all NLIs is 11 m. The white arrows indicate the round islands in each case.}\label{JSF-un}
\end{figure}

\begin{figure}[htb]
	\includegraphics[width=9cm]{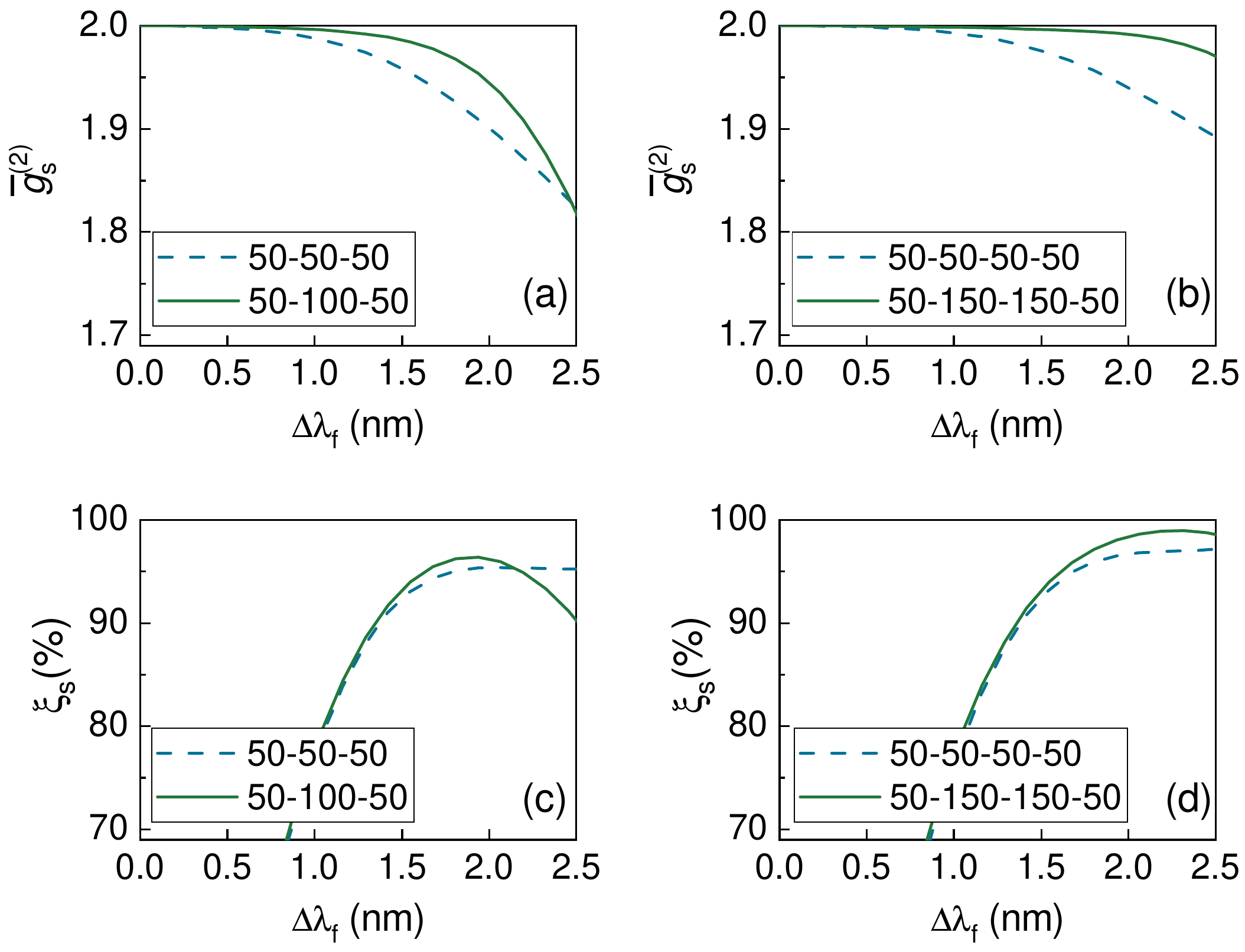}
	\caption{Comparison of $\bar g_{s}^{(2)}$ [(a), (b)] and  heralding efficiency $\xi_{s}$ [(c), (d)] between uneven (solid curves) and even (dashed curves) length for 3-stage [(a), (c)] and 4-stage [(b), (d)] NLIs, respectively.}\label{eff-g2}
\end{figure}

To demonstrate the advantage of this approach, we run simulations for 3-stage and 4-stage NLIs with uneven DSF length distribution of $50-100-50$ and $50-150-150-50$, respectively, and compare the results to their even length counterparts (using Eqs. (\ref{multi-JSF}, \ref{multi-H})). The numbers here denote the length of individual DSFs in the unit of meter. The length of individual SMFs is 11 m for all cases. Fig.\ref{JSF-un} shows the contour plot of the JSF for these two cases (a, b) as compared to their even length counterparts of 50-50-50 and 50-50-50-50 (c, d). It can be seen that there is no mini maxima in Figs. \ref{JSF-un}(a, b). For the effectiveness of the arrangement, we next evaluate the intensity correlation function and collection efficiency for the round islands of each case, which are indicated by the white arrows in Fig. \ref{JSF-un}. Due to the similarity for the results in signal and idler fields, we only display the calculated results of $\bar g_{s}^{(2)}$ and $\xi_{s}$ as a function of the filter bandwidth of the signal channel, as shown in Fig.\ref{eff-g2}. As can be seen, the uneven cases (solid curves) significantly improve upon the even cases (dashed curves).

\section{High gain regime}

So far, our discussion is limited to the JSF, which directly characterizes the frequency correlation between signal and idler photons generated in spontaneous parametric process. This is in the low gain regime of operation for the parametric processes, that is, the parameter $G$ in Eq.(\ref{TPS}) is much smaller than one. This corresponds to the case of low pump power. But when the pump power is high so that $G\sim 1$ or even much larger than 1, the parametric process becomes an amplification process with appreciable gain. The parametric process in this case can produce entangled states with EPR-type correlations in continuous variables \cite{guo-OE16, Guo16OL}. However, the output quantum state cannot be obtained from Eq.(\ref{TPS}) but must be derived from
\begin{equation}\label{QS}
|\Psi\rangle = \hat U |vac\rangle,
\end{equation}
via the evolution operator:
\begin{eqnarray}\label{U}
\hat U = \exp\left\{\frac{1}{ i\hbar}\int dt \hat H\right\}
\end{eqnarray}
with \cite{ou-multi, sil}
\begin{eqnarray}\label{H}
\int dt\hat H   = i\hbar G \int d \omega_1 d\omega_2F_{NLI}(\omega_1,\omega_2)\hat a_s ^{\dag}(\omega_1)\hat a_i^{\dag}(\omega_2)+ h.c.,~~~~
\end{eqnarray}
where $F_{NLI}(\omega_1,\omega_2)$ is the same JSF in Eq.(\ref{TPS}) but is now modified by the NLI. Note that the states in Eqs.(\ref{TPS}) and (\ref{TFPS}) are the first order and second order approximation of Eq.(\ref{QS}) when $G\ll 1$.

However, for large $G$, it is better to work in the Heisenberg picture where the input and output operators are connected by
\begin{eqnarray}
&&\hat{b}_s(\omega_s)= \hat{U}^{\dag} \hat{a}_s(\omega_s) \hat{U} ,\cr
&&\hat{b}_i(\omega_i)= \hat{U}^{\dag} \hat{a}_i(\omega_i) \hat{U},
\end{eqnarray}
and it can be shown that for the evolution operator given in Eq.(\ref{U}), the output operators are related to the input by a set of Green functions: \cite{Guo13PRA}
\begin{subequations}
\label{multimode_trans}
\begin{equation}
\hat{b}_s(\omega_s)=\hat{U}^{\dag} \hat{a}_s(\omega_s) \hat{U}=\int_S h_{1s}(\omega_s,\omega_s')\hat{a}_s(\omega_s') d\omega_s'+ \int_I h_{2s}(\omega_s,\omega_i')\hat{a}_i^{\dag}(\omega_i') d\omega_i'
\end{equation}
\begin{equation}
\hat{b}_i(\omega_i)=\hat{U}^{\dag} \hat{a}_i(\omega_i) \hat{U}=\int_I h_{1i}(\omega_i,\omega_i')\hat{a}_i(\omega_i') d\omega_i'+ \int_S h_{2i}(\omega_i,\omega_s')\hat{a}_s^{\dag}(\omega_s') d\omega_s' ,
\end{equation}
\end{subequations}
with the Green functions given by
\begin{equation}\label{h1s}
\begin{aligned}
h_{1s}(\omega_s,\omega_s') = & \delta(\omega_s-\omega_s') + \sum_{n=1}^{\infty} \frac{G^{2n}}{(2n)!} \iint \cdots \int d\omega_1 d\omega_2 \cdots d\omega_{2n-1}\\
\bigg\lbrace & \big[ F_{NLI}(\omega_s,\omega_1) F_{NLI}(\omega_2,\omega_3) \dots F_{NLI}(\omega_{2n-2},\omega_{2n-1}^{ }) \big] \\
& \big[ F_{NLI}^* (\omega_2,\omega_1) F_{NLI}^* (\omega_4,\omega_3) F_{NLI}^* (\omega_6,\omega_5) \dots F_{NLI}^ *(\omega_{s}',\omega_{2n-1} ) \big] \bigg\rbrace,
\end{aligned}
\end{equation}
\begin{equation}\label{h2s}
\begin{aligned}
h_{2s}(\omega_s,\omega_i') = & G F_{NLI}(\omega_s,\omega_i') + \sum_{n=1}^{\infty} \frac{G^{2n+1}}{(2n+1)!} \iint \cdots \int d\omega_1 d\omega_2 \cdots d\omega_{2n}\\
\bigg\lbrace & \big[ F_{NLI}^*(\omega_2,\omega_1) F_{NLI}^*(\omega_4,\omega_3) \dots F_{NLI}^*(\omega_{2n},\omega_{2n-1}^{ }) \big] \\
& \big[ F_{NLI} (\omega_s,\omega_1) F_{NLI} (\omega_2,\omega_3) F_{NLI} (\omega_4,\omega_5) \dots F_{NLI}^*(\omega_{2n},\omega_i^{'} ) \big] \bigg\rbrace,
\end{aligned}
\end{equation}
\begin{equation}\label{h1i}
\begin{aligned}
h_{1i}(\omega_i,\omega_i') = & \delta(\omega_i-\omega_i') + \sum_{n=1}^{\infty} \frac{G^{2n}}{(2n)!} \iint \cdots \int d\omega_1 d\omega_2 \cdots d\omega_{2n-1}\\
\bigg\lbrace & \big[ F_{NLI}(\omega_1,\omega_i) F_{NLI}(\omega_3,\omega_2) \dots F_{NLI}(\omega_{2n-1},\omega_{2n-2}^{ }) \big] \\
& \big[ F_{NLI}^* (\omega_1,\omega_2) F_{NLI}^* (\omega_3,\omega_4) F_{NLI}^* (\omega_5,\omega_6) \dots F_{NLI}^ *(\omega_{2n-1},\omega_{i}' ) \big] \bigg\rbrace,
\end{aligned}
\end{equation}
\begin{equation}\label{h2i}
\begin{aligned}
h_{2i}(\omega_i,\omega_s') = & G F_{NLI}(\omega_s',\omega_i) + \sum_{n=1}^{\infty} \frac{G^{2n+1}}{(2n+1)!} \iint \cdots \int d\omega_1 d\omega_2 \cdots d\omega_{2n}\\
\bigg\lbrace & \big[ F_{NLI}^*(\omega_1,\omega_2) F_{NLI}^*(\omega_3,\omega_4) \dots F_{NLI}^*(\omega_{2n-1},\omega_{2n}^{ }) \big] \\
& \big[ F_{NLI} (\omega_1,\omega_i) F_{NLI} (\omega_3,\omega_2) F_{NLI} (\omega_5,\omega_4) \dots F_{NLI}(\omega_s',\omega_{2n} ) \big] \bigg\rbrace.
\end{aligned}
\end{equation}
It has been shown that the spectral property of the signal and idler photons is directly related to the Green function $h_{2s(i)}(\omega_{s(i)},\omega_{i(s)})$ and both $h_{2s}(\omega_s,\omega_i)$ and $h_{2i}(\omega_i,\omega_s)$ have the same pattern \cite{Guo13PRA}. So, we only numerically calculate $h_{2s}(\omega_s,\omega_i)$  as a function of the gain coefficient $G$ in order to illustrate the change of the spectral property influenced by gain parameter.
Note that for $G\ll 1$, $h_{2s}(\omega_s,\omega_i) \propto F_{NLI}(\omega_s,\omega_i)$. The detailed parameters of $F_{NLI} (\omega_s,\omega_i)$ are the same as that listed for Fig.\ref{p-JSF} in Sect.~III-C. For large $G$, we use Eq.(\ref{h2s}) up to 40 terms to calculate $h_{2s}(\omega_s,\omega_i)$, since our simulation shows that the accuracy does not change after 40 terms.

\begin{figure}
	\includegraphics[width=14cm]{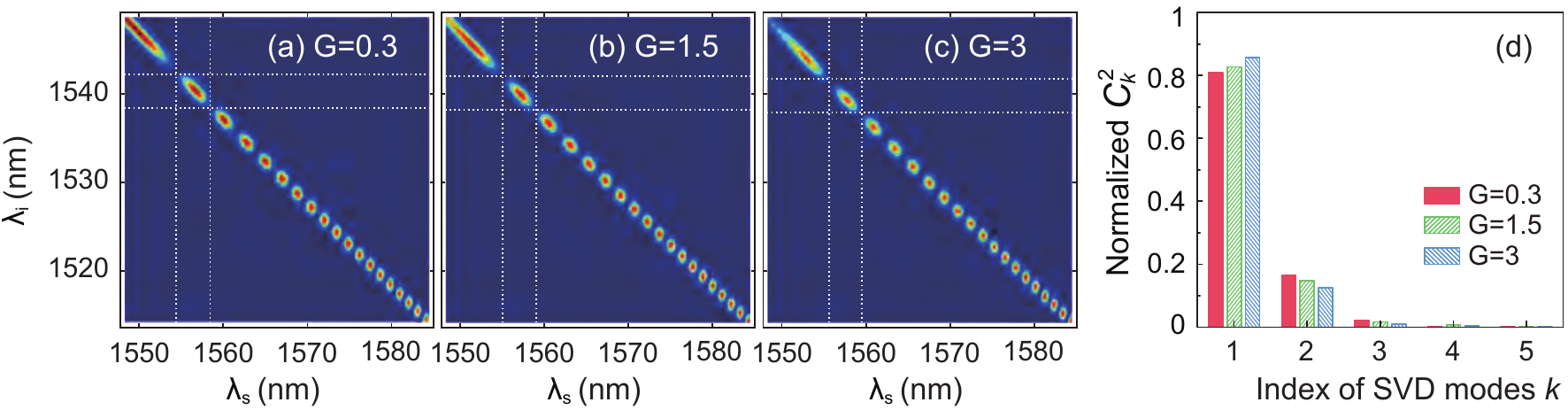}
	\caption{ Contour plot of the absolute square of $h_{2s}(\omega_s,\omega_i)$ as a function of the gain parameter $G$ with (a) $G=0.3$, (b) $G=1.5$, (c) $G=3$. (d) Mode indice of SVD of $h_{2s}(\omega_s,\omega_i)$. The dashed lines illustrate the filters of bandwidth 3.6 nm.
	}\label{High gain}
\end{figure}

The contour plots of the Green function $h_{2s(i)}(\omega_{s(i)},\omega_{i(s)})$ are shown in Fig.~\ref{High gain}(a)-(c) with $G = 0.3,~1.5,~3$, respectively.
As can be seen, the Green functions maintain the quasi-periodically island pattern with the gain parameter $G$ increased, but the islands seem to be getting more rounded. So, the technique of NLI can be used to engineer the spectrum of parametric processes in the high gain regime as well \cite{Huo19arx}.

In order to reveal the mode properties in more detail, we can use filters $f_s^{(m)}(\omega_s), f_i^{(m)}(\omega_i)$ to select out the specific island, just like what we did in Sect.III-D. For $m=1$ island, we use filters of bandwidth 3.6 nm and make a single-value decomposition of the normalized function of the filtered $h_{2s}$
under different gain condition. We show the mode index coefficients in Fig.~\ref{High gain}(d). Apparently, the lowest order coefficient will become larger as $G$ increases. This explains the trend shown in Fig.~\ref{High gain}(a)-(c) that the islands seem to be getting more rounded as $G$ increases and is consistent with the general trend of the parametric process at large gain, that is, the first order mode dominates and the process becomes near single-mode operation as $G$ increases \cite{LiuNN-OE16}.
We experimentally demonstrated this feature for the single selected island in the NLI-modified spectrum of the parametric process \cite{Huo19arx}.  Of course, the better way to make the islands round is to adjust the pump bandwidth and the length of the SMF, as discussed before in Eq.(\ref{pump-length}).

\section{Summary and Discussion}

In this paper, we analyze a novel interferometric method to engineer the joint spectral function (JSF) of a two-photon state generated from spontaneous parametric emission. We achieve this by employing nonlinear interferometer (NLI) schemes. We successfully modify an original frequency anti-correlated JSF from dispersion shifted fiber to a nearly factorized JSF using a two-stage NLI. We further refine the two-stage NLI with a programmable optical filter and extend our discussion to a multi-stage NLI for finer engineering.

The role played by NLI in JSF engineering is spectral filtering. But different from passive filtering after the production of the photon pairs, which may destroy the correlation between the photons in a pair, the spectral filtering achieved in NLI is an active filtering scheme that selectively produces the photon pair and thus preserves the photon correlation. Because of this, the engineered two-photon states maintain a high collection efficiency while having the spectral properties modified to become frequency uncorrelated for near unity modal purity, which gives rise to a transform-limited two-photon state.

Our investigation provides a new approach for engineering the spectral property of photon pairs and for obtaining narrow band photon pairs simultaneously possessing the advantages of high purity, high collection efficiency and high brightness, which are an important resource of quantum states for quantum information and communication. Compared with the methods of directly generating factorable photon pairs by engineering the dispersion property of nonlinear medium, our NLI approach is easy to implement and has flexible output. The wavelength of photon pairs can be conveniently tuned for various tasks. Compared with the methods of generating factorable photon pairs by applying narrow band filters in signal and idler bands, our NLI approach does not decrease the brightness of photon pairs and collection efficiency. Moreover, by introducing the multi-stage design into our NLI approach, we are able to develop a source delivering fully factorable photon pairs entangled in multi-frequency channels for high dimensional entanglement.

The interferometric approach discussed here can also be applied to high gain situation for tailoring the spectrum, which is even more critical in exploring quantum entanglement with continuous variables \cite{tatia, Chekhova16, Sharapova18PRA, guo-OE16, Fabre19REV, Huo19arx}. However, the case becomes complicated when the pump power is increased in nonlinear fibers for high gain because power-dependent nonlinear phase shift will alter the spectral properties of the generated fields as well. Losses is also a critical factor to consider in quantum entanglement with continuous variables in the design involving a programmable optical filter and multiple stages. Nevertheless, the interferometric approach provides us more flexibility in engineering quantum states.

\begin{acknowledgments}
This work was supported in part by the National Natural Science Foundation of China (Nos. 11527808, 91736105), and the National Key Research and Development Program of China (No. 2016YFA0301403), and by US National Science Foundation (No.1806425).

\end{acknowledgments}

\nocite{*}
\bibliographystyle{apsrev4-1}
\bibliography{NLI-P-PRA-APR20}

\appendix*
\section{Derivation of the heralded auto-intensity correlation function}
In this appendix we give the procedure for deriving Eqs. (\ref{HSPSf4}) to (\ref{Hg2}), which finally lead to the expression of the heralded auto-intensity correlation function $\tilde g_s^{(2)}$.
The state given in Eq.(\ref{TPS}) is a low pump power approximation for the quantum state of the light field from  spontaneous parametric processes. Higher photon number terms will start to contribute when the pump power is high in order to increase the brightness of the source. The next order modification includes a four-photon state and leads to a state of \cite{ou99}
\begin{eqnarray}\label{TFPS-A}
|\Psi\rangle \approx |vac\rangle + G|\Psi_2\rangle + (G^2/2) |\Psi_4\rangle
\end{eqnarray}
with $|\Psi_2\rangle$ given in Eq.(\ref{TPS2}) and the four-photon modification of
\begin{eqnarray}\label{FPS-A}
|\Psi_4\rangle &= &|\Psi_2\rangle\otimes |\Psi_2\rangle\cr
&=&\int d \omega_s d\omega_id \omega_s' d\omega_i'F(\omega_s,\omega_i)F(\omega_s',\omega_i')\hat a_s ^{\dag}(\omega_s)\hat a_s ^{\dag}(\omega_s')\hat a_i^{\dag}(\omega_i)\hat a_i^{\dag}(\omega_i')|vac\rangle.
\end{eqnarray}
The four-photon term corresponds to two-pair production and will lead to a two-photon state in the heralded field. We will calculate this two-photon state from the four-photon modification term in Eq.(\ref{FPS-A}) in this Appendix.

For completeness of discussion, we will include the passive filters. As in Eq.(\ref{TPSf}), the passive filters are modeled as beam splitters by replacing $\hat a_s(\omega)$ with $\hat a_s'\equiv f_s(\omega)\hat a_s(\omega)+r_s(\omega)\hat a_{sv}(\omega)$ and $\hat a_i(\omega)$ with $\hat a_i'\equiv f_i(\omega)\hat a_i(\omega)+r_i(\omega)\hat a_{iv}(\omega)$. The contributions to the heralded state from the first two terms in Eq.(\ref{TFPS-A}) are exactly the same as Eq.(\ref{HSPSf2}). So, we will only calculate the contribution from the four-photon term here.

The heralding detection of an idler photon at time $t$ collapses the four-photon state in Eq.(\ref{FPS-A}) to
\begin{eqnarray}\label{HS-A}
\hat E_i(t)|\bar\Psi_4\rangle
&=&\int d \omega_s d\omega_id \omega_s' d\omega_i'F(\omega_s,\omega_i)F(\omega_s',\omega_i')\hat a_s ^{'\dag}(\omega_s)\hat a_s ^{'\dag}(\omega_s') \hat E_i(t) \hat a_i^{'\dag}(\omega_i)\hat a_i^{'\dag}(\omega_i')|vac\rangle\cr
&=&\frac{1}{ \sqrt{2\pi}}\int d \omega_s d\omega_id \omega_s' d\omega_i'F(\omega_s,\omega_i)F(\omega_s',\omega_i')\hat a_s ^{'\dag}(\omega_s)\hat a_s ^{'\dag}(\omega_s') \cr
&&\hskip 1in \times\Big[e^{-j\omega_i t}\hat a_i^{'\dag}(\omega_i')f_i(\omega_i) +e^{-j\omega_i' t}\hat a_i^{'\dag}(\omega_i)f_i(\omega_i')\Big]|vac\rangle,
\end{eqnarray}
where $|\bar\Psi_4\rangle$ is the filtered state and $\hat E_i(t) = \frac{1}{ \sqrt{2\pi}}\int d\omega e^{-j\omega t}\hat a_i(\omega)$. With slow heralding detectors unable to time resolve the pulsed field, the projected density operator becomes
\begin{eqnarray}\label{Proj-A}
\hat {\bar \rho}_{proj} &=& \frac{G^4}{ 4} \int dt \hat E_i(t)|\bar\Psi_4\rangle \langle\bar\Psi_4|\hat E_i^{\dag}(t)\cr
&=&\frac{G^4}{ 4}\int d \omega_s d \omega_s' d \bar \omega_s d\bar \omega_s' |1_s'(\omega_s)1_s'(\omega_s')\rangle\langle 1_s'(\bar \omega_s)1_s'(\bar \omega_s')|\cr
 &&\hskip 0.5in \times \int d\omega_id\omega_i'd\bar \omega_id\bar \omega_i'F(\omega_s,\omega_i)F(\omega_s',\omega_i')
F^*(\bar \omega_s,\bar \omega_i)F^*(\bar \omega_s',\bar \omega_i')\cr
&&\hskip 0.7in \times \frac{1}{2\pi}\int dt \Big[e^{-j\omega_i t}f_i(\omega_i)|1_i'(\omega_i')\rangle +e^{-j\omega_i' t}f_i(\omega_i')|1_i'(\omega_i)\rangle\Big]\cr
&&\hskip 1.28in \times \Big[e^{j\bar\omega_i t}f_i(\bar\omega_i)\langle 1_i'(\bar\omega_i')| +e^{j\bar\omega_i' t}f_i(\bar\omega_i')\langle1_i'(\bar\omega_i)|\Big].
\end{eqnarray}
After carrying out the time integral and trace out the idler photons, we obtain
\begin{eqnarray}\label{2pden-A}
\hat {\bar\rho}_2 &=& {\rm Tr}_i \hat {\bar \rho}_{proj} \cr
&=&\frac{G^4}{4} \int d \omega_s d \omega_s' d \bar \omega_s d\bar \omega_s' |1_s'(\omega_s)1_s'(\omega_s')\rangle\langle 1_s'(\bar \omega_s)1_s'(\bar \omega_s')|\cr
 &&\hskip 0.5in \times \int d\omega_id\omega_i'F(\omega_s,\omega_i)F(\omega_s',\omega_i') [f_i^2(\omega_i)+ f_i^2(\omega_i')]\cr
&&\hskip 0.7in \times [F^*(\bar \omega_s,\omega_i)F^*(\bar \omega_s',\omega_i')+
F^*(\bar \omega_s,\omega_i')F^*(\bar \omega_s',\omega_i)].
\end{eqnarray}
Proper normalization requires the evaluation of the trace of the density operator, which gives
\begin{eqnarray}\label{Tr-bar-rho2}
{\rm Tr} \hat {\bar\rho}_2 &=&  G^4 \int d \omega_s d \omega_s' d\omega_id\omega_i' f_i^2(\omega_i) F(\omega_s,\omega_i)F(\omega_s',\omega_i')\cr &&\hskip 0.3in \times \big[ F^*(\omega_s,\omega_i)F^*(\omega_s',\omega_i')+ F^*(\omega_s,\omega_i')F^*(\omega_s',\omega_i)\big] \cr
&=& G^4(\bar{\cal A}_i + \bar{\cal E}_i),
\end{eqnarray}
where $\bar{\cal A}_i$, $\bar{\cal E}_i$ are similar to those in Eq.(\ref{EA}) except the filtering factor $f_i^2(\omega_i)$ is included.
Tracing out the vacuum photons in the signal field, we obtain the two-photon part of the heralded state in the signal field upon detection of one idler photon:
\begin{eqnarray}\label{bar-rho2}
\hat {\bar\rho}_2' =  \hat {\tilde \rho}_0  + \hat {\tilde \rho}_1 + \hat {\tilde \rho}_2,
\end{eqnarray}
where $\hat {\tilde \rho}_0  , \hat {\tilde \rho}_1$ are the vacuum and one-photon terms whose exact forms are unimportant because they only give higher order corrections to the state in Eq.(\ref{HSPSf2}). $\hat {\tilde \rho}_2$ are the two-photon term and has the form of
\begin{eqnarray}\label{tilt-rho2}
&&\hat {\tilde \rho}_2 = \frac{G^4}{4} \int d \omega_s d \omega_s' d\bar\omega_s d\bar\omega_s' f_s(\omega_s)f_s(\omega_s')f_s(\bar \omega_s)f_s(\bar \omega_s') \cr &&\hskip 0.3in \times \int d\omega_id\omega_i'F(\omega_s,\omega_i)F(\omega_s',\omega_i')\big[ F^*(\bar\omega_s,\omega_i)F^*(\bar\omega_s',\omega_i')+ F^*(\bar\omega_s,\omega_i')F^*(\bar\omega_s',\omega_i)\big] \cr &&\hskip 0.3in \times
\big[f_i^2(\omega_i)+f_i^2(\omega_i')\big]|1_s(\omega_s)1_s(\omega_s')\rangle \langle 1_s(\bar\omega_s)1_s(\bar\omega_s')|.
\end{eqnarray}

Combining the above with Eq.(\ref{HSPSf2}) which is the contribution from the first two terms of Eq.(\ref{TFPS-A}),
we find the normalized heralded state as
\begin{eqnarray}\label{HSPSf4-A}
{\hat {\bar \rho}}''
=  {\cal N} \bigg[T \sum_k \bar r_k^2 |\bar 1_k\rangle\langle \bar 1_k| + R|vac\rangle\langle vac| + G^2 (\hat {\tilde \rho}_0'  + \hat {\tilde \rho}_1' + \hat {\tilde \rho}_2')/4\bar {\cal A}_i^{1/2}\bigg],
\end{eqnarray}
where $\hat {\tilde \rho}_l'\equiv 4 \hat {\tilde \rho}_l/ G^4 $ for $l=0,1,2$ with $\bar {\cal A}_i$ given in Eq.(\ref{HSPSf2-Tr}). ${\cal N} = [1+ C G^2]^{-1}$ is the normalization factor with $C$ as some constant related to the JSF $F(\omega_s,\omega_i)$ and filter functions $f_s,f_i,r_s,r_i$. ${\cal N} \approx 1$ for $G\ll1$.

The heralded auto-intensity correlation function $\tilde g_s^{(2)}$ is defined as
\begin{eqnarray}
\tilde g_s^{(2)}\equiv \frac{\int dt_1dt_2 \tilde \Gamma_s^{(2)}(t_1,t_2)}{
\big[\int dt\tilde \Gamma_s^{(1)}(t)\big]^2}\label{h-g2}
\end{eqnarray}
with
\begin{eqnarray}
\tilde \Gamma_s^{(2)}(t_1,t_2) \equiv  {\rm Tr}[{\hat {\bar \rho}}''\hat I_s(t_1)\hat I_s(t_2)],~~~~
\tilde \Gamma_s^{(1)}(t)  \equiv {\rm Tr}[{\hat {\bar \rho}}''\hat I_s(t)] .\label{h-Gamma}
\end{eqnarray}
$\hat {\tilde \rho}_2'$ is the only term in $\hat {\bar \rho}''$ that will contribute to $\tilde \Gamma_s^{(2)}(t_1,t_2)$, which is calculated as
\begin{equation}
\begin{aligned}
{\tilde \Gamma_s^{(2)}}({t_1},{t_2})
& = {\rm{Tr}}\left[ {\hat{\bar{ \rho}} ''{{\hat I}_s}({t_1}){{\hat I}_s}({t_2})} \right]\\
& = \frac{1}{(2\pi)^2}\frac{{{G^2}}}{{4{{\bar {\cal A}}^{1/2}}_i}}{\rm{Tr}}\left[ {{{\hat{\tilde{\rho}} '}_2}\int {d{{\tilde \omega }_{s1}}d{{\tilde \omega '}_{s1}}d{{\tilde \omega }_{s2}}d{{\tilde \omega '}_{s2}}{e^{ - {\rm{i(}}{{\tilde \omega }_{s1}}{t_1}{\rm{ + }}{{\tilde \omega }_{s2}}{t_2})}}} } \right. \\
&\quad\quad \left.{ \times {e^{{\rm{i(}}{{\tilde \omega '}_{s1}}{t_{\rm{1}}}{\rm{ + }}{{\tilde \omega '}_{s2}}{t_2})}}\hat a_s^\dag ({{\tilde \omega '}_{s1}})\hat a_s^\dag ({{\tilde \omega '}_{s2}}){{\hat a}_s}({{\tilde \omega }_{s2}}){{\hat a}_s}({{\tilde \omega }_{s1}})} \right ]\\
& = \frac{1}{(2\pi)^2}\frac{{{G^2}}}{{4{{\bar {\cal A}}^{1/2}}_i}}\int {d{\omega_s}d{{\omega_s '}}d{{\bar \omega }_s}d{{\bar \omega_s '}}} {f_s}({\omega_s}){f_s}({{\omega '}_s}){f_s}({{\bar \omega }_s}){f_s}({{\bar \omega '}_s})\\
& \quad\quad \times \int {d{\omega _i}d{{\omega_i '}}} F({\omega _s},{\omega _i})F({{\omega_s '}},{{\omega_i '}})\left[ {f_i^2({\omega_i}) + f_i^2({{\omega_i '}})} \right]\\
& \quad\quad \times \left[ {{F^*}({{\bar \omega }_s},{\omega_i}){F^*}({{\bar \omega '}_s},{{\omega '}_i}) + {F^*}({{\bar \omega }_s},{{\omega_i '}}){F^*}({{\bar \omega_s '}},{\omega _i})} \right]\\
& \quad\quad \times \left[ {{e^{-{\rm{i(}}{\omega_s}{t_1} + {{\omega_s '}}{t_2})}} \! +\! {e^{-{\rm{i(}}{\omega_s}{t_{\rm{2}}} + {{\omega_s '}}{t_{\rm{1}}})}}} \right]\left[ {{e^{{\rm{i(}}{{\bar \omega }_s}{t_1} + {{\bar \omega '}_s}{t_2})}} \! +\! {e^{{\rm{i(}}{{\bar \omega }_s}{t_{\rm{2}}} + {{\bar \omega_s '}}{t_{\rm{1}}})}}} \right]. \\
\end{aligned}
\end{equation}
Then we carry out the time integral
\begin{equation}
\begin{aligned}\label{Gamma-s-2}
\int {d{t_1}d{t_2}{{\tilde \Gamma_s }^{(2)}}({t_1},{t_2})}
& = \frac{{{G^2}}}{{4{{\bar {\cal A}}^{1/2}}_i}}\int {d{\omega_s}d{{\omega_s '}}d{{\bar \omega }_s}d{{\bar \omega_s '}}} {f_s}({\omega _s}){f_s}({{\omega_s '}}){f_s}({{\bar \omega }_s}){f_s}({{\bar \omega_s '}})\\
& \quad \times \int {d{\omega _i}d{{\omega_i '}}} F({\omega_s},{\omega _i})F({{\omega_s '}},{{\omega_i '}})\left[ {f_i^2({\omega_i}) + f_i^2({{\omega_i '}})} \right]\\
& \quad \times \left[ {{F^*}({{\bar \omega_s }},{\omega_i}){F^*}({{\bar \omega_s '}},{{\omega_i '}}) + {F^*}({{\bar \omega_s }},{{\omega_i '}}){F^*}({{\bar \omega_s'}},{\omega_i})} \right]\\
& \quad \times 2\left[ {\delta ({\omega _s} - {{\bar \omega }_s})\delta ({{\omega_s '}} - {{\bar \omega_s '}}) + \delta ({{\omega_s '}} - {{\bar \omega }_s})\delta ({\omega_s} - {{\bar \omega_s '}})} \right]\\
& {\rm{ = }}\frac{{{\rm{2}}{P_s}{P_c}}}{{{P_i}}}(1 + \frac{{\bar {\cal E}}}{{\bar {\cal A}}}),
\end{aligned}
\end{equation}
with
\begin{equation}
\bar {\cal A} = \int {d{\omega _s}d{{\omega '}_s}} \int {d{\omega _i}d{{\omega '}_i}} f_s^2({\omega _s})f_s^2({\omega '_s})f_i^2({\omega _i}){\left| {F({\omega _s},{\omega _i})F({{\omega '}_s},{{\omega '}_i})} \right|^2}
\end{equation}
and
\begin{equation}
\begin{aligned}
\bar {\cal E} = \int {d{\omega _s}d{{\omega '}_s}} & \int {d{\omega _i}d{{\omega '}_i}} f_s^2({\omega _s})f_s^2({{\omega '}_s})f_i^2({\omega _i})\\
& \times F({\omega _s},{\omega _i}){F^*}({\omega _s},{{\omega '}_i})F({{\omega '}_s},{{\omega '}_i}){F^*}({{\omega '}_s},{\omega _i}),
\end{aligned}
\end{equation}
where $P_c,P_s,P_i$ are given in Eq.(\ref{EQ_Psi}, \ref{P_c}) with $\eta_s=1=\eta_i$.
From Eq.(\ref{HSPSf2}), we find
\begin{equation}
\begin{aligned}
{{\tilde \Gamma_s }^{({\rm{1}})}}(t) & = {\rm{Tr}}\left[ {\hat{\bar{ \rho}} ''\hat I(t)} \right]\\
& = {\rm{Tr}}\left[ {{\hat{\bar{\rho}} '}_1}\hat{E}_s^{\dagger}(t)\hat{E}_s(t) \right]\\
& = \frac{1}{2\pi }\frac{1}{P_i}{\rm{Tr}}\left[ \int d{\omega_s}d{{\omega_s '}} d{\omega_i}F({\omega_s},{\omega_i}){F^*}(\omega_s ',\omega_i) f_i^2({\omega_i}){f_s}({\omega_s}){f_s}(\omega_s ')e^{ - {\rm{i}}({\omega_s} - {\omega_s} ')t} \right],
\end{aligned}
\end{equation}
then we carry out the time integral
\begin{equation}
\begin{aligned}\label{Gamma-s-1}
\int {dt} {\tilde \Gamma_s }^{(\rm{1})}(t)
& = \frac{1}{2\pi} \frac{1}{P_i} \int dt \rm{Tr} \left[ \int {d{\omega _s}d{{\omega_s '}}} d{\omega _i}F({\omega_s},{\omega_i}){F^*}({{\omega_s '}},{\omega_i})\right. \\
& \left. \quad \times {f_i^2({\omega_i}){f_s}({\omega_s}){f_s}({{\omega_s '}}){e^{ - {\rm{i(}}{\omega_s} - {{\omega_s '}})t}}} \right]\\
& = \frac{1}{P_i} \int {d{\omega _s}} d{\omega _i}{\left| {F({\omega _s},{\omega _i})} \right|^2}f_s^2({\omega _s})f_i^2({\omega _i})\\
& = \frac{{{P_c}}}{{{P_i}}}.
\end{aligned}
\end{equation}
After substituting Eqs.(\ref{Gamma-s-2}) and (\ref{Gamma-s-1}) into Eq.(\ref{h-g2}), we arrive the heralded auto-intensity correlation function
\begin{equation}
\tilde g_s^{(2)} = \frac{{2{P_c}}}{{{h_i}{h_s}}}(1 + \frac{{\bar {\cal E}}}{{\bar {\cal A}}}),
\end{equation}
where $h_i,h_s$ are the heralding efficiencies given in Eq.(\ref{collection}) with $\eta_i=1=\eta_s$.

\end{document}